%% file: icml2022_arxiv.tex
\pdfoutput=1

\documentclass[nohyperref]{article}

\input{math_commands.tex}

\usepackage{lipsum}
\usepackage{url}
\usepackage{subcaption}
\usepackage{multicol}
\usepackage{multirow}
\usepackage{siunitx}
\usepackage{wrapfig}
\usepackage{amssymb}

\usepackage{microtype}
\usepackage{graphicx}
\usepackage{booktabs} %

\usepackage{xcolor}
\usepackage{tikz}
\usetikzlibrary{bayesnet}
\usetikzlibrary{arrows}
\usetikzlibrary{shapes.multipart}
\usetikzlibrary{positioning}
\tikzset{
state/.style={
       rectangle split,
       rectangle split parts=2,
       rectangle split part fill={red!30,blue!20},
       rounded corners,
       draw=black, very thick,
       minimum height=2em,
       text width=3cm,
       inner sep=2pt,
       text centered,
       }
}
\include{bayesnet_extras}

\definecolor{customgreen} {RGB}{217	234	212}
\definecolor{customblue}  {RGB}{205	226	242}
\definecolor{customorange}{RGB}{254	228	207}
\definecolor{customred}{RGB}{217 144 144}
\definecolor{custommorered}{RGB}{218 46 42}
\definecolor{customred}{RGB}{255 182 173}%
\definecolor{sharedcolour}{RGB}{254	228	207}

\colorlet{observed-color}{customgreen}
\colorlet{latent-color}{customblue}
\colorlet{deterministic-color}{gray!15}
\colorlet{deterministic-skip-color}{custommorered}
\colorlet{shared-color}{blue}

\usepackage{hyperref}

\usepackage[accepted]{icml2022_arxiv}

\usepackage[capitalize,noabbrev]{cleveref}

\renewcommand\paragraph[1]{\textbf{#1.}\hspace{1ex}}

\icmltitlerunning{Benchmarking Generative Latent Variable Models for Speech}

\begin{document}

\twocolumn[
    \icmltitle{Benchmarking Generative Latent Variable Models for Speech}
    
    \icmlsetsymbol{equal}{*}
    \begin{icmlauthorlist}
    \icmlauthor{Jakob D. Havtorn}{corti,dtu}
    \icmlauthor{Lasse Borgholt}{corti,ku}
    \icmlauthor{S\o ren Hauberg}{dtu}
    \icmlauthor{Jes Frellsen}{dtu}
    \icmlauthor{Lars Maal\o e}{corti,dtu}
    \end{icmlauthorlist}
    
    \icmlaffiliation{corti}{Corti.ai, Copenhagen, Denmark}
    \icmlaffiliation{dtu}{Department of Applied Mathematics and Computer Science, Technical University of Denmark, Kongens Lyngby, Denmark}
    \icmlaffiliation{ku}{Department of Computer Science, University of Copenhagen, Copenhagen, Denmark}
    
    \icmlcorrespondingauthor{Jakob D. Havtorn}{jdh@corti.ai}
    
    \icmlkeywords{latent variable models, variational inference, sequence models, speech, audio, generative models, probabilistic modeling}
    
    \vskip 0.3in
]

\printAffiliationsAndNotice{}  %

\begin{abstract}
Stochastic latent variable models (LVMs) achieve state-of-the-art performance on natural image generation but are still inferior to deterministic models on speech. In this paper, we develop a speech benchmark of popular temporal LVMs and compare them against state-of-the-art deterministic models. We report the likelihood, which is a much used metric in the image domain, but rarely, or incomparably, reported for speech models. To assess the quality of the learned representations, we also compare their usefulness for phoneme recognition. Finally, we adapt the Clockwork VAE, a state-of-the-art temporal LVM for video generation, to the speech domain. Despite being autoregressive only in latent space, we find that the Clockwork VAE can outperform previous LVMs and reduce the gap to deterministic models by using a hierarchy of latent variables.
\end{abstract}

\section{Introduction}
After the introduction of the variational autoencoder (VAE,  \citealp{kingma_auto-encoding_2014, rezende_stochastic_2014}) quickly came two temporal extensions for modeling speech data \citep{chung_recurrent_2015, fraccaro_sequential_2016}. Since then, temporal LVMs have undergone little development compared to their counterparts in the image domain, where LVMs recently showed superior performance to autoregressive models such as PixelCNN \citep{oord_pixel_2016, oord_conditional_2016, salimans_pixelcnn_2017}. The improvements in the image domain have been driven mainly by top-down inference models and deeper latent hierarchies \citep{sonderby_ladder_2016, maaloe_biva_2019, vahdat_nvae_2020, child_very_2021,sinha_consistency_2021,kingma_variational_2021}. In speech modeling however, autoregressive models such as the WaveNet remain state-of-the-art \citep{oord_wavenet_2016}.

To compare and develop LVMs for speech, we need good benchmarks similar to those in the image domain. Image benchmarks commonly compare likelihood scores, but research in the speech domain often omits reporting a likelihood \citep{oord_wavenet_2016, hsu_unsupervised_2017, oord_neural_2018} or report likelihoods that are incomparable due to subtle differences in the assumed data distribution \citep{chung_recurrent_2015, fraccaro_sequential_2016, hsu_unsupervised_2017, aksan_stcn_2019}. Without a proper comparison standard, it is difficult for the field of explicit likelihood models on speech to develop in an informed manner.

To advance the state of LVMs for speech, this paper (i) develops a benchmark for LVMs based on model likelihood, (ii) introduces a hierarchical LVM architecture without autoregressive decoder, (iii) compares LVMs to deterministic counterparts including WaveNet, and (iv) qualitatively and quantitatively evaluates the latent variables learned by different LVMs based on their usefulness for phoneme recognition. We find that:
\begin{itemize}\vspace{-3mm}
    \setlength\itemsep{-0em}
    \item [(I)] State-of-the-art LVMs achieve likelihoods that are inferior to WaveNet at high temporal resolution but are superior at lower resolutions. Interestingly, we find that a standard LSTM \citep{hochreiter_long_1997} almost matches the likelihood of WaveNet.
    \item [(II)] LVMs with powerful autoregressive decoders achieve better likelihoods than the non-autoregressive LVM.
    \item[(III)] The expressiveness of LVMs for speech increases with a deeper hierarchy of stochastic latent variables, similar to conclusions within image modeling. %
    \item[(IV)] LVMs learn rich representations that are as good or better than Mel spectrograms for phoneme recognition also when using only 10 minutes of labeled data.
\end{itemize}\vspace{-3mm}
At a high level, this benchmark brings order to LVM model comparisons for speech and also provides useful reference implementations of the models\footnote{\fontsize{8.5}{0}\selectfont\texttt{ \href{https://github.com/JakobHavtorn/benchmarking-lvms}{github.com/JakobHavtorn/benchmarking-lvms}}}. Before presenting the results, we provide a brief survey of existing LVMs for speech in a coherent notation.

\section{Latent variable models for speech}
\begin{figure*}[t!]
    \centering
    \begin{multicols}{5}
    \resizebox{!}{6cm}{
    \begin{tikzpicture}
        \node[obs] (x_t) {$\vx_t$};
        
        \node[above=0.4cm of x_t] (caption) {LSTM};
        
        \node[det,below=2.55cm of x_t] (d_t) {$\vd_{t}$};
        \node[det,left=0.75cm of d_t] (d_tm1) {$\vd_{t-1}$};
        \node[obs,below=0.75cm of d_t] (x_tm1) {$\vx_{t-1}$};
        
        \edge[deterministic-skip-color]{x_tm1}{d_t};
        \edge[deterministic-skip-color]{d_t}{x_t};
        \edge[deterministic-skip-color]{d_tm1}{d_t};
        \edge[draw=white,bend right]{d_t}{x_t};  %
    \end{tikzpicture}
    }
    \\
    \resizebox{!}{6cm}{
    \begin{tikzpicture}
        \node[obs] (x_t) {$\vx_t$};
        
        \node[above=0.4cm of x_t] (caption) {VRNN};
        
        \node[latent,below=0.75cm of x_t] (z_t) {$\vz_t$};
        \node[latent,left=0.75cm of z_t] (z_tm1) {$\vz_{t-1}$};
        
        \node[det,below=0.75cm of z_t] (d_t) {$\vd_{t}$};
        \node[det,below=0.75cm of z_tm1] (d_tm1) {$\vd_{t-1}$};
        \node[obs,below=0.75cm of d_t] (x_tm1) {$\vx_{t-1}$};
        
        \edge[deterministic-skip-color]{x_tm1}{d_t};
        \edge[]{d_t}{z_t};
        \edge[]{z_tm1}{d_t};
        \edge[deterministic-skip-color]{d_tm1}{d_t};
        \edge[]{z_t}{x_t};
        \edge[deterministic-skip-color,bend right]{d_t}{x_t};
    \end{tikzpicture}
    }
    \\
    \resizebox{!}{6cm}{
    \begin{tikzpicture}
        \node[obs] (x_t) {$\vx_t$};
        
        \node[above=0.4cm of x_t] (caption) {SRNN};
        
        \node[latent,below=0.75cm of x_t] (z_t) {$\vz_t$};
        \node[latent,left=0.75cm of z_t] (z_tm1) {$\vz_{t-1}$};
        
        \node[det,below=0.75cm of z_t] (d_t) {$\vd_{t}$};
        \node[det,below=0.75cm of z_tm1] (d_tm1) {$\vd_{t-1}$};
        \node[obs,below=0.75cm of d_t] (x_tm1) {$\vx_{t-1}$};
        
        \edge[deterministic-skip-color]{x_tm1}{d_t};
        \edge[]{d_t}{z_t};
        \edge[]{z_tm1}{z_t};
        \edge[deterministic-skip-color]{d_tm1}{d_t};
        \edge[]{z_t}{x_t};
        \edge[deterministic-skip-color,bend right]{d_t}{x_t};
    \end{tikzpicture}
    }
    \resizebox{!}{6cm}{
    \begin{tikzpicture}
        \node[obs] (x_t) {$\vx_t$};
        \node[above=0.4cm of x_t] (caption) {STCN};
        
        \node[latent,below=0.75cm of x_t] (z_t) {$\vz_t$};
        \node[left=0.025cm of z_t] (dots) {$\cdots$};
        \node[latent,left=0.75cm of z_t] (z_tmr) {$\vz_{t-r'}$};
        
        \node[det,below=0.75cm of z_t] (d_t) {$\vd_{t}$};
        \node[obs,below=0.75cm of d_t] (x_tm1) {$\vx_{t-1}$};
        \node[left=0.025cm of x_tm1] (dots) {$\cdots$};
        \node[obs,left=0.75cm of x_tm1] (x_tmr) {$\vx_{t-r}$};
        
        \edge[]{x_tm1}{d_t};
        \edge[]{x_tmr}{d_t};
        \edge[]{d_t}{z_t};
        \edge[]{z_t}{x_t};
        \edge[]{z_tmr}{x_t};
        \edge[draw=white,bend right]{d_t}{x_t};  %
    \end{tikzpicture}
    }
    \\
    \resizebox{!}{4.73cm}{
    \begin{tikzpicture}
        \node[obs] (x_t) {$\vx_t$};%

        \node[above=0.4cm of x_t] (caption) {CW-VAE};
        
        \node[latent,below=0.75cm of x_t] (z_t) {$\vz_t$};
        \node[latent,left=0.75cm of z_t] (z_tm1) {$\vz_{t-1}$};
        
        \node[det,below=0.75cm of z_t] (d_t) {$\vd_{t}$};
        \node[det,below=0.75cm of z_tm1] (d_tm1) {$\vd_{t-1}$};
        
        \edge[]{d_t}{z_t};
        \edge[bend right]{d_t}{x_t};
        \edge[]{z_tm1}{d_t};
        \edge[]{d_tm1}{d_t};
        \edge[]{z_t}{x_t};
    \end{tikzpicture}
    }
    \end{multicols}
    \vspace{-0.5cm}
\caption{
Generative models for a single time step of a deterministic autoregressive LSTM, the VRNN and SRNN as well as the STCN and CW-VAE both with a single layer of latent variables. 
Red arrows indicate purely deterministic paths from the output $\vx_t$ to previous input $\vx_{<t}$ without passing a stochastic node. 
In an LSTM, information flows deterministically from previous observed variables to the next. 
The VRNN and SRNN use stochastic latent variables but also include deterministic skip connections from previous observed variables. 
The STCN is also autoregressive in $\vx$ but does not use deterministic skip connections from $\vx_{t-1}$ to $\vx_t$. 
The CW-VAE is not autoregressive on the observed variable which forces information to flow through the latent variables. 
We provide more elaborate graphical illustrations including inference models of the CW-VAE in \figref{fig: clockwork vae inference generative models 2 layered} and of the VRNN, SRNN and STCN in appendix~\ref{app: additional graphical models}.
}
\label{fig: autoregressive vrnn srnn cwvae 1L graphs}
\vspace{-2mm}
\end{figure*}
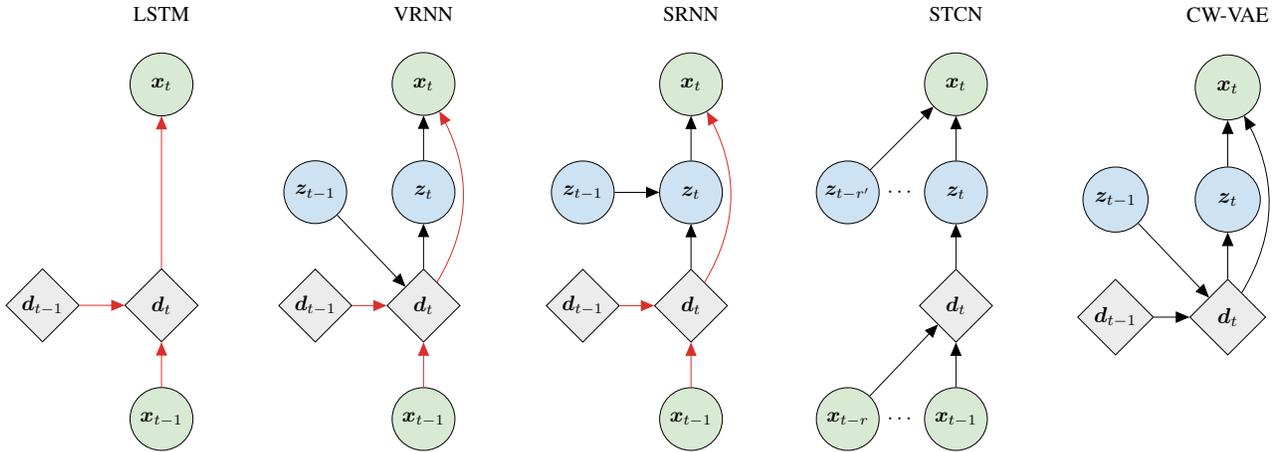

LVMs formulated via the VAE framework continue to be of interest due to their ability to learn an approximation to the posterior distribution of assumed latent variables. The posterior is usually of a significantly reduced dimensionality compared to the input and lies close to a known prior distribution. Approximate posteriors are useful for tasks beyond generation such as semi-supervised learning \citep{kingma_semi-supervised_2014} and anomaly detection \citep{havtorn_hierarchical_2021}.

In recent years, several complementary methods have been proposed to improve the expressiveness of VAEs. These include building more expressive priors via methods such as normalizing flows \citep{rezende_variational_2015} and building a deeper hierarchy of stochastic latent variables such as the Ladder VAE \citep{sonderby_ladder_2016}. In this research, we focus on the latter due to the recent breakthroughs in image modeling using VAEs without costly autoregressive dependencies on the observed variable \citep{maaloe_biva_2019,vahdat_nvae_2020,child_very_2021}.

Several works have applied LVMs to speech. Among the first contributions were the VRNN \cite{chung_recurrent_2015} and SRNN \citep{fraccaro_sequential_2016} which can be seen as conditional VAEs per timestep. Other recent LVMs include the FH-VAE \citep{hsu_unsupervised_2017}, which leverages an additional latent variable to capture global features, and Z-forcing \cite{goyal_z-forcing_2017}, which resembles the SRNN but includes an auxiliary task in the latent space to increase its utilization. The VQ-VAE \citep{oord_neural_2018} is a hybrid between an LVM and an autoregressive model which uses a quantized latent space to improve the quality of generated samples. The Stochastic WaveNet \cite{lai_stochastic_2018} and STCN \cite{aksan_stcn_2019} use WaveNet encoder and decoders and temporally independent latent variables.

In this paper, we focus on the VRNN, SRNN and STCN. We exclude the Stochastic WaveNet as it is similar to STCN and achieves inferior likelihoods \cite{aksan_stcn_2019}. The FH-VAE, with disjoint latent variables and discriminative objective, Z-forcing, with an auxiliary task, and the VQ-VAE, with a quantized latent space and autoregressive prior fitted after training, all introduce significant changes to the original VAE framework and are also not included here.

All selected models have autoregressive generative models which let future observed variables be generated by conditioning on previously generated values. Inspired by recent progress in the image domain, we therefore formulate and benchmark a novel temporal LVM which does not rely on an autoregressive decoder. We do so by adapting the hierarchical Clockwork Variational Autoencoder \citep{saxena_clockwork_2021}, originally proposed for video generation, to speech.

\subsection{Sequential deep latent variable models}
The selected models are all sequential deep latent variable models trained with variational inference and the reparameterization trick \cite{kingma_auto-encoding_2014}.
They take as input a variable-length sequence $\vx_{1:T} = \left(\vx_1, \vx_2, \dots, \vx_{T}\right)$ with $\vx_t \in \mathbb{R}^{D_x}$. 
We let $\vx_{1:T}$ refer to the observed variable or a downsampled version of it. We will sometimes use $\vx$ to refer to the sequence $\vx_{1:T}$ when it is not ambiguous.

First, $\vx_{1:T}$ is encoded to a temporal stochastic latent representation $\vz_{1:T} = \left( \vz_1, \vz_2, \dots, \vz_{T} \right)$ with $\vz_t\in \mathbb{R}^{D_z}$. This representation is then used to reconstruct the original input $\vx_{1:T}$.
The latent variable is assumed to follow some prior distribution $p(\vz_t|\cdot)$ where the dot indicates that it may depend on latent and observed variables at previous time steps, $\vz_{<t}$ and $\vx_{<t}$ where $\vz_{<t} := \left( \vz_1, \vz_2, \dots, \vz_{t-1} \right)$. 

The models are trained to maximize a likelihood objective. The exact marginal likelihood is given by
\begin{equation}
    \log p_\vtheta(\vx) = \log \int p_\vtheta(\vx,\vz) \, \mathrm{d}\vz  \enspace,
\end{equation}
where $\vtheta$ are parameters of the generative model. The exact likelihood is intractable to optimize due to the integration over the latent space. Instead, we introduce the variational approximation $q_\vphi(\vz|\vx)$ to the true posterior. Via Jensen's inequality this yields the well-known evidence lower bound (ELBO) on the exact likelihood 
\begin{align}
    \log p_\vtheta(\vx) 
    &\geq \mathbb{E}_{q_\vphi(\vz|\vx)} \left[ \log \frac{p_\vtheta(\vx,\vz)}{q_\vphi(\vz|\vx)} \right] \, \mathrm{d}\vz =: \mathcal{L}(\vtheta,\vphi;\vx) \label{eq: elbo}
\end{align}
which can be jointly optimized with respect to  $\{\vtheta, \vphi\}$ using stochastic gradient descent methods. We omit the $\vtheta$ and $\vphi$ subscripts for the remainder of the paper.
A graphical illustration of the models can be seen in \figref{fig: autoregressive vrnn srnn cwvae 1L graphs}. Additional graphical models can be found in appendix~\ref{app: additional graphical models}.

\subsection{Variational recurrent neural network ({\small VRNN})}
The VRNN \citep{chung_recurrent_2015} is essentially a VAE per timestep. At timestep $t$, the VAE is conditioned on the hidden state of a recurrent neural network (RNN), $\vd_{t}\in\mathbb{R}^{D_d}$, with state transition $\vd_{t} = f([\vx_{t-1}, \vz_{t-1}], \vd_{t-1})$ where $[\cdot, \cdot]$ denotes concatenation. 
The VRNN uses a Gated Recurrent Unit (GRU, \citealp{cho_properties_2014}) for $f$.
The joint distribution factorizes over time and the latent variables are autoregressive in both the observed and latent space,
\begin{equation}
    p(\vx_{1:T},\vz_{1:T}) = \prod_{t=1}^{T} p(\vx_t|\vx_{<t},\vz_{\leq t}) p(\vz_t|\vx_{<t},\vz_{<t}) \enspace . \label{eq: vrnn joint factorization}
\end{equation}
The approximate posterior similarly factorizes over time,
\begin{equation}
    q(\vz_{1:T}|\vx_{1:T}) = \prod_{t=1}^{T} q(\vz_t|\vx_{\leq t},\vz_{<t}) \enspace . \label{eq: vrnn inference factorization}
\end{equation}
From this, the ELBO for the VRNN is
\begin{equation}
\begin{split}
    \mathcal{L}(\vx) =&~ \mathbb{E}_{q(\vz_{1:T}|\vx_{1:T})} \bigg[ \sum_{t} \log p(\vx_t|\vx_{<t},\vz_{\leq t}) \\ & - \text{KL}\left( q(\vz_t|\vx_{\leq t},\vz_{<t}) \, || \, p(\vz_t|\vx_{<t},\vz_{<t}) \right)  \bigg] \enspace .
    \label{eq: vrnn elbo}
\end{split}
\end{equation}
The VRNN uses diagonal covariance Gaussian distributions $\mathcal{N}$ for the prior and posterior distributions. We denote the output distribution of choice by $\mathcal{D}$. 
\begin{align}
    q(\vz_t|\vx_{\leq t},\vz_{<t}) &= \mathcal{N}\left(\valpha_{q}(\vx_{t},\vd_{t})\right) \label{eq: vrnn parameterization inference} \\
    p(\vz_t|\vx_{<t},\vz_{<t}) &= \mathcal{N}\left(\valpha_{p}(\vd_t)\right) \label{eq: vrnn parameterization prior} \\
    p(\vx_t|\vx_{<t},\vz_{\leq t}) &= \mathcal{D}\left(\vbeta(\vz_{t},\vd_{t})\right) \enspace . \label{eq: vrnn parameterization observation}
\end{align}
All sets of distributional parameters, $\valpha_q,\valpha_p$ and $\vbeta$, are the outputs of densely connected neural networks which we notationally overload as functions in equations~(\ref{eq: vrnn parameterization inference}-\ref{eq: vrnn parameterization observation}). It is common to refer to $\valpha_q$ as the inference model or encoder and $\vbeta$ as the decoder. Together with $\vbeta$, the structural model $\valpha_p$ forms the generative model.

Since the decoder is dependent on $\vd_t$, the transition function $f$ allows the VRNN to learn to ignore parts of or the entire latent variable and establish a purely deterministic transition from $\vx_{t-1}$ to $\vd_t$ (\figref{fig: autoregressive vrnn srnn cwvae 1L graphs}). This failure mode is commonly referred to as posterior collapse and is a well-known weakness of VAEs with powerful decoders \cite{bowman_generating_2016, sonderby_ladder_2016}.

\subsection{Stochastic recurrent neural network ({\small SRNN})}
The SRNN \citep{fraccaro_sequential_2016} is similar to the VRNN but differs by separating the stochastic latent variables from the deterministic representations (\figref{fig: autoregressive vrnn srnn cwvae 1L graphs}). That is, the GRU state transition is independent of $\vz_{1:T}$ such that $\vd_{t} = f(\vx_{t-1}, \vd_{t-1})$. With this, the joint $p(\vx_{1:T},\vz_{1:T})$ can be written as for the VRNN in \eqref{eq: vrnn joint factorization}. The approximate posterior of $\vz_t$ is conditioned on the full observed sequence,
\begin{align}
    \label{eq: srnn inference factorization}
    q(\vz_{1:T}|\vx_{1:T}) = \prod_{t=1}^{T} q(\vz_t|\vx_{1:T},\vz_{t-1}) \enspace .
\end{align}
This is achieved by introducing a second GRU that runs backwards in time with transition $\va_t = g([\vx_t,\vd_t], \va_{t+1})$. While $p(\vx_t|\vx_{<t},\vz_{\leq t})$ remains as in \eqref{eq: vrnn parameterization observation}, we have
\begin{align}
    q(\vz_t|\vx_{1:T},\vz_{<t}) &= \mathcal{N}\left(\valpha_{q}(\vz_{t-1},\va_{t})\right) \label{eq: srnn parameterization inference} \\
    p(\vz_t|\vx_{<t},\vz_{<t}) &= \mathcal{N}\left(\valpha_{p}(\vz_{t-1},\vd_t)\right) \enspace . \label{eq: srnn parameterization prior}
\end{align}
By inferring $\vz_t$ conditioned on the full sequence $\vx_{1:T}$, the SRNN performs smoothing. This has been noted to better approximate the true posterior of $\vz_t$ which can be shown to depend on the full observed sequence \cite{bayer_mind_2021}. Comparatively, the VRNN performs filtering.

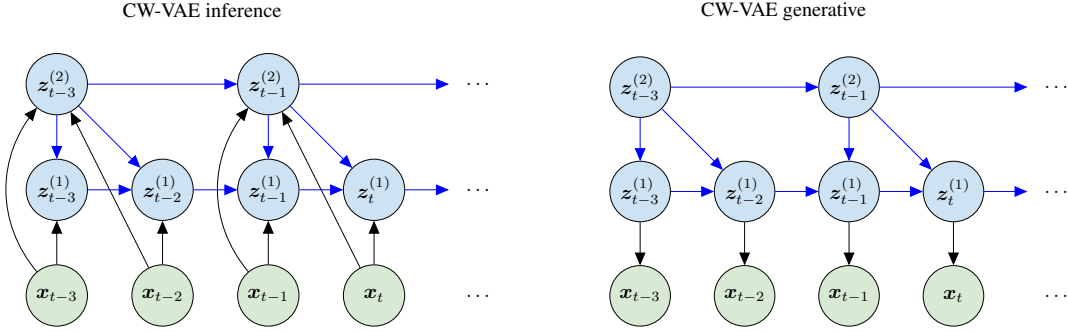
\begin{figure*}[t]
    \centering
    \resizebox{!}{4.4cm}{
    \begin{tikzpicture}
        \node[obs] (x_1) {$\vx_{t-3}$};
        \node[obs,right=.75cm of x_1] (x_2) {$\vx_{t-2}$};
        \node[obs,right=.75cm of x_2] (x_3) {$\vx_{t-1}$};
        \node[obs,right=.75cm of x_3] (x_4) {$\vx_{t}$};
        \node[obs,right=.75cm of x_4,fill=white,draw=white] (x_5) {$\dots$};

        \node[latent,above=.75cm of x_1](s_1_1){$\vz^{(1)}_{t-3}$};
        \node[latent,above=.75cm of x_2](s_1_2){$\vz^{(1)}_{t-2}$};
        \node[latent,above=.75cm of x_3](s_1_3){$\vz^{(1)}_{t-1}$};
        \node[latent,above=.75cm of x_4](s_1_4){$\vz^{(1)}_{t}$};
        \node[latent,above=.75cm of x_5,fill=white,draw=white](s_1_5){$\dots$};

        \node[latent,above=.75cm of s_1_1](s_2_1){$\vz^{(2)}_{t-3}$};
        \node[latent,above=.75cm of s_1_3](s_2_3){$\vz^{(2)}_{t-1}$};
        \node[latent,above=.75cm of s_1_5,fill=white,draw=white](s_2_5){$\dots$};

        \edge[]{x_1}{s_1_1};
        \edge[]{x_2}{s_1_2};
        \edge[]{x_3}{s_1_3};
        \edge[]{x_4}{s_1_4};

        \edge[bend left=40]{x_1}{s_2_1};
        \edge[]{x_2}{s_2_1};
        \edge[bend left=40]{x_3}{s_2_3};
        \edge[]{x_4}{s_2_3};

        \edge[shared-color]{s_1_1}{s_1_2};
        \edge[shared-color]{s_1_2}{s_1_3};
        \edge[shared-color]{s_1_3}{s_1_4};
        \edge[shared-color]{s_1_4}{s_1_5};
        
        \edge[shared-color]{s_2_1}{s_2_3};
        \edge[shared-color]{s_2_3}{s_2_5};
        
        \edge[shared-color]{s_2_1}{s_1_1};
        \edge[shared-color]{s_2_3}{s_1_3};

        \edge[shared-color]{s_2_1}{s_1_2};
        \edge[shared-color]{s_2_3}{s_1_4};
        
        \node[above=of s_2_1, yshift=-0.5cm, xshift=2.5cm] (phi) {CW-VAE inference};
    \end{tikzpicture}
    }
    \hspace{1cm}
    \resizebox{!}{4.4cm}{
    \begin{tikzpicture}
        \node[obs] (x_1) {$\vx_{t-3}$};
        \node[obs,right=.75cm of x_1] (x_2) {$\vx_{t-2}$};
        \node[obs,right=.75cm of x_2] (x_3) {$\vx_{t-1}$};
        \node[obs,right=.75cm of x_3] (x_4) {$\vx_{t}$};
        \node[obs,right=.75cm of x_4,fill=white,draw=white] (x_5) {$\dots$};

        \node[latent,above=.75cm of x_1](s_1_1){$\vz^{(1)}_{t-3}$};
        \node[latent,above=.75cm of x_2](s_1_2){$\vz^{(1)}_{t-2}$};
        \node[latent,above=.75cm of x_3](s_1_3){$\vz^{(1)}_{t-1}$};
        \node[latent,above=.75cm of x_4](s_1_4){$\vz^{(1)}_{t}$};
        \node[latent,above=.75cm of x_5,fill=white,draw=white](s_1_5){$\dots$};

        \node[latent,above=.75cm of s_1_1](s_2_1){$\vz^{(2)}_{t-3}$};
        \node[latent,above=.75cm of s_1_3](s_2_3){$\vz^{(2)}_{t-1}$};
        \node[latent,above=.75cm of s_1_5,fill=white,draw=white](s_2_5){$\dots$};

        \edge[]{s_1_1}{x_1};
        \edge[]{s_1_2}{x_2};
        \edge[]{s_1_3}{x_3};
        \edge[]{s_1_4}{x_4};

        \edge[shared-color]{s_1_1}{s_1_2};
        \edge[shared-color]{s_1_2}{s_1_3};
        \edge[shared-color]{s_1_3}{s_1_4};
        \edge[shared-color]{s_1_4}{s_1_5};
        
        \edge[shared-color]{s_2_1}{s_2_3};
        \edge[shared-color]{s_2_3}{s_2_5};
        
        \edge[shared-color]{s_2_1}{s_1_1};
        \edge[shared-color]{s_2_3}{s_1_3};
        
        \edge[shared-color]{s_2_1}{s_1_2};
        \edge[shared-color]{s_2_3}{s_1_4};

        \node[above=of s_2_1, yshift=-0.5cm, xshift=2.5cm] (phi) {CW-VAE generative};
    \end{tikzpicture}
    }
\caption{
Inference (left) and generative (right) models for the CW-VAE with a hierarchy of $L=2$ latent variables, $s_1=1$ and $s_2=2$. 
The models are unrolled over four consecutive time steps but note that the graph continues towards $t=0$ and $t=T$. 
Blue arrows indicate parameter sharing between inference and generative models. 
We omit the deterministic variable of \figref{fig: autoregressive vrnn srnn cwvae 1L graphs} for clarity.
}
\label{fig: clockwork vae inference generative models 2 layered}
\vspace{-0mm}
\end{figure*}

\subsection{Stochastic temporal convolutional network ({\small STCN})}
The STCN \cite{aksan_stcn_2019} is a hierarchical latent variable model with an autoregressive generative model based on WaveNet \cite{oord_wavenet_2016}. Contrary to VRNN and SRNN, the latent variables are independent in the sense that there are no transition functions connecting them over time.
Instead, a latent $\vz_t^{(l)}$ at layer $l$ is conditioned on a window of observed variables $\vx_{\scriptscriptstyle\mathcal{R}_t^{(l)}}$ via a WaveNet encoder. The windowed dependency introduced by the WaveNet is defined via the index set $\mathcal{R}_t^{(l)} = \cbra{t-r_l+1,\dots,t}$ where $r_l$ is the receptive field of the encoder at layer $l$. The window size grows exponentially with the layer $l$. The joint can be written as
\begin{equation}
    p(\vx_{1:T},\vz_{1:T}) = \prod_{t=1}^{T} p(\vx_t|\vz^{(1:L)}_{{\scriptscriptstyle \mathcal{R}}_{t}^{(1)}}) \prod_{l=1}^{L} p(\vz^{(l)}_{t}|\vx_{\scriptscriptstyle\mathcal{R}_{t-1}^{(l)}},\vz^{(l+1)}_{t}) \nonumber
    \label{eq: stcn joint factorization}
\end{equation}
where $\vz_t^{(L+1)}:=\emptyset$ for notational convenience. 
The deterministic encoding is $\vd_t^{(l)} = h(\vx_{\scriptscriptstyle\mathcal{R}_{t}^{(l)}})$ where $h$ is the encoder and $\vd_t^{(l)}$ is extracted from the $l$'th layer similar to a Ladder VAE \cite{sonderby_ladder_2016}. 
The approximate posterior is
\begin{equation}
    q\pa{\vz_{1:T}^{(1:L)}|\vx_{1:T}} = \prod_{t=1}^{T}\prod_{l=1}^{L} q\pa{\vz_{t}^{(l)}|\vx_{\scriptscriptstyle\mathcal{R}_{t}^{(l)}},\vz_t^{(l+1)}}
\end{equation}
The factorized distributions are given as
\begin{align}
    q\left(\vz_t^{(l)}|\vx_{\scriptscriptstyle\mathcal{R}_{t}^{(l)}},\vz_t^{(l+1)}\right) &= \mathcal{N}\pa{\valpha_{q}^{(l)}\pa{\vz_t^{(l+1)},\vd_t^{(l)}}} \label{eq: stcn parameterization inference} \\
    p\left(\vz_t^{(l)}|\vx_{\scriptscriptstyle\mathcal{R}_{t-1}^{(l)}},\vz_{t}^{(l+1)}\right) &= \mathcal{N}\pa{\valpha_{p}^{(l)}\pa{\vz_t^{(l+1)},\vd_{t-1}^{(l)}}} \label{eq: stcn parameterization prior} \\
    p\left(\vx_t|\vz^{(1:L)}_{\scriptscriptstyle\mathcal{R}_{t}^{(1)}}\right) &= \mathcal{D}\pa{\vbeta\pa{\vz^{(1:L)}_{\scriptscriptstyle\mathcal{R}_{t}^{(1)}}}} \enspace . \label{eq: stcn parameterization observation}
\end{align}
The decoder $\vbeta(\vz^{(1:L)}_{\scriptscriptstyle\mathcal{R}_{t}^{(1)}})$ is also a WaveNet.

\subsection{Clockwork variational autoencoder ({\small CW-VAE})}
The CW-VAE \citep{saxena_clockwork_2021} is a hierarchical latent variable model recently introduced for video generation. As illustrated in figures~\ref{fig: autoregressive vrnn srnn cwvae 1L graphs} and \ref{fig: clockwork vae inference generative models 2 layered}, it is autoregressive in the latent space but not in the observed space, contrary to the VRNN, SRNN and STCN. Additionally, each latent variable is updated only every $s_l$ timesteps, where $s_l$ is a layer-dependent integer, or stride, and $s_1<s_2<\dots<s_L$.
This imposes the inductive bias that latent variables exist at different temporal resolutions with $\vz^{(l)}$ changing at lower frequency than $\vz^{(l-1)}$. 
In speech, phonetic variation between $\SI{10}{}-\SI{400}{ms}$, morphological and semantic features at the word level and speaker-related variation at the global scale make this a reasonable assumption.

To simplify temporal notation, we define the timesteps at which a layer updates its latent state as $\mathcal{T}_l := \{t \in \bra{1,T} \,|\, (t-1)\,\text{mod}\, s_l = 0\}$. We then define the set of layers that update at a given timestep as $\mathcal{J}_t := \{ l \,|\, t\in\mathcal{T}_l \}$. The joint distribution can then be written as,
\begin{equation}
    p(\vx_{1:T},\vz_{1:T}^{(1:L)}) = 
    \prod_{t=1}^{T} p(\vx_t | \vz_t^{(1)})
    \prod\limits_{l\in\mathcal{J}_t} p(\vz_t^{(l)} | \vz_{t-1}^{(l)}, \vz_t^{(l+1)}) \enspace . \nonumber
    \label{eq: cwvae joint}
\end{equation}
The inference model is conditioned on a window of the observed variable $\vx_{t:t+s_l}$ depending on the layer's stride $s_l$.
\begin{equation}
    q\pa{\vz_{1:T} | \vx_{1:T}} = \prod_{t=1}^T \prod\limits_{l\in\mathcal{J}_t} q\pa{\vz_t^{(l)} | \vz_{t-1}^{(l)}, \vz_t^{(l+1)}, \vx_{t:t+s_l}} \enspace . \nonumber
    \label{eq: cwvae inference}
\end{equation}
The dependency on $\vx_{t:t+s_l}$ is encoded via a convolutional ladder network similar to the STCN with $\vd_t^{(l)} = e(\vx_{t:t+s_l})$. We define $\vx_{\scriptscriptstyle\mathcal{S}_t^l}:=\vx_{t:t+s_l}$ for compactness. The latent state transitions are densely connected and the decoder is also a convolutional network. 
\begin{align}
    q\left(\vz_t^{(l)}|\vx_{\scriptscriptstyle\mathcal{S}_t^l},\vz_{t-1}^{(l)},\vz_t^{(l+1)}\right) &= \mathcal{N}\pa{\valpha_{q}^{l}\pa{\vz_{t-1}^{(l)},\vz_{t}^{(l+1)},\vd_t^{(l)}}} \nonumber \\ %
    p\left(\vz_t^{(l)}|\vz_{t-1}^{(l)},\vz_{t}^{(l+1)}\right) &= \mathcal{N}\pa{\valpha_{p}^{l}\pa{\vz_{t-1}^{(l)},\vz_t^{(l+1)}}} \nonumber \\
    p\left(\vx_t|\vz^{(1)}_{t-s_l/2:t+s_l/2}\right) &= \mathcal{D}\pa{\vbeta\pa{\vz^{(1)}_{t-s_l/2:t+s_l/2}}} \nonumber %
\end{align}

\subsection{Speech modeling with Clockwork VAEs} \label{sec: cwvae on speech}
The video and speech modalities differ in the sampling rates normally used to digitize the natural signals. Sampling rates of common video codecs typically range from $\SI{24}{Hz}$ up to $\SI{60}{}$ or $\SI{120}{Hz}$. In the speech domain, sampling rates range from $\SI{8000}{Hz}$ up to e.g.\@ $\SI{44100}{Hz}$ commonly used for music recordings. In the original work, $s_l$ is defined as $s_l:= k^{l-1}$ for some constant $k$ which makes it exponentially dependent on the layer index $l$ and forces $s_1 = 1$. While this is reasonable for the sample rates of video, training a model at this resolution is infeasible for audio waveform modeling. For this reason, we chose $s_1 \gg 1$ to achieve an initial temporal downsampling and let $s_l:= c^{l-1}s_1$ for $l>1$ and some constant $c$.

The encoder and decoder of the original CW-VAE are not applicable to speech. Hence, we parameterize them using 1D convolutions operate on the raw waveform. We use a ladder-network, similar to \citet{sonderby_ladder_2016, aksan_stcn_2019}, for the encoder as this provides benefits compared to alternatives such as a standalone encoder per latent variable. 
Specifically, a ladder-network leverages parameter sharing across the latent hierarchy and importantly processes the full observed sequence only once sharing the resulting representations between latent variables. This yields a more computationally efficient encoder and higher activity in latent variables towards the top of the hierarchy.

\subsection{Output distribution}
Audio, as well as image data, are naturally continuous signals that are represented in discrete form in computers. The signals are sampled with some bit depth $b$ which defines the range of attainable values, $\vx\in\{0,1,\dots,2^b-1\}$. The bit depth typically used in audio and image samplers is between $\SI{8}{}$ and $\SI{32}{bit}$ with $\SI{8}{bit}$ and $\SI{16}{bit}$ being the most common in the literature (MNIST, \citealt{lecun_gradient-based_1998}; CelebA, \citealt{liu_deep_2015}; CIFAR10, \citealt{krizhevsky_learning_2009}; TIMIT, \citealt{garofolo_timit_1993}; LibriSpeech, \citealt{panayotov_librispeech_2015}).

In the image domain, the discrete nature of the data is usually modeled in one of two ways; either by using discrete distributions \cite{salimans_pixelcnn_2017, maaloe_biva_2019, child_very_2021} or by dequantizing the data and using a continuous distribution \cite{dinh_nice_2015, sonderby_ladder_2016, ho_flow_2019}, which yields a lower bound on the discrete distribution likelihood \cite{theis_note_2016}. 
In the speech domain, however, the output distribution is often taken to be a continuous Gaussian \citep{hsu_unsupervised_2017,lai_stochastic_2018,zhu_s3vae_2020} which was also originally done in VRNN, SRNN and STCN. 
This choice generally results in an ill-posed problem with a likelihood that is unbounded from above unless the variance is lower bounded \citep{mattei_leveraging_2018}. As a result, \emph{reported likelihoods can be sensitive to hyperparameter settings and be hard to compare}. We discuss this phenomenon further in appendix~\ref{app: gaussian likelihood unboundedness discussion}.

Most work normalizes the audio or standardizes it to values in a bounded interval $\vx\in[-1,1]$. Since $\vx$ becomes approximately continuous as the bit depth $b$ becomes large and the range of possible values becomes small, this alleviates the issue. However, commonly used datasets with bit-depths of $\SI{16}{}$ still result in a discretization gap between values that remains much larger than the gap between the almost continuous 32 bit floating point numbers which reinforces the problem \citep{bishop_pattern_2006}.

In this work, we therefore benchmark models using a discretized mixture of logistics (DMoL) as output distribution. The DMoL was introduced for image modeling with autoregressive models \citep{salimans_pixelcnn_2017} but has become standard in other generative models \citep{maaloe_biva_2019, vahdat_nvae_2020, child_very_2021}. 
It was recently applied to autoregressive speech modeling of raw waveforms \citep{oord_parallel_2018}. 
As opposed to e.g.\@ a categorical distribution, the DMoL induces ordinality on the observed space such that values that are numerically close are also close in a probabilistic sense. This is a sensible inductive bias for images as well as audio where individual samples represent the amplitude of light or pressure, respectively. We discuss the DMoL for audio further in appendix~\ref{app: additional discussion on output distribution}.

\section{Speech modeling benchmark}
\paragraph{Data}
We train models on TIMIT \citep{garofolo_timit_1993}, LibriSpeech \citep{panayotov_librispeech_2015} and LibriLight \citep{kahn_libri-light_2020}. For TIMIT, we randomly sample 5\% of the training split for validation. 
We represent the audio as $\mu$-law encoded PCM standardized to values in $\bra{-1, 1}$ with discretization gap of ${2^{-b+1}}$. We use the original bit depth of $\SI{16}{bits}$ and sample rate of $\SI{16000}{Hz}$. We use this representation both as the input and the reconstruction target. We provide more details on the datasets in appendix~\ref{app: dataset details} and additional results on linear PCM in appendix~\ref{app: additional likelihood results appendix}.

\paragraph{Likelihood}
We report likelihoods in units of bits per frame (bpf) as this yields a more interpretable and comparable likelihood than total likelihood in nats. It also has direct connections with information theory and compression \citep{shannon_mathematical_1948, townsend_practical_2019}. In units of bits per frame, lower is better. For LVMs, we report the one-sample ELBO. The likelihoods can be seen in \twotabref{tab: likelihoods timit}{tab: likelihoods librispeech}. We describe how to convert likelihood to bpf in appendix~\ref{app: likelihood in bits per frame}.

\paragraph{Models}
Architecture and training details are sketched below, while the full details are in appendices~\ref{app: model architectures} and \ref{app: training details} along with additional results for some alternative model configurations in appendix~\ref{app: additional likelihood results appendix}.
We select model configurations that can be trained on GPUs with a maximum of 12GB of RAM and train all models until convergence on the validation set.
We use a DMoL with 10 components for the output distribution of all models and model all datasets at their full bit depth of $\SI{16}{bits}$. We train and evaluate models on stacked waveforms similar to previous work \cite{chung_recurrent_2015, fraccaro_sequential_2016, aksan_stcn_2019} with stack sizes of $s=1$, $s=64$ and $s=256$. Hence, every model input $\vx_t$ is composed of $\tilde{x}_{t:t+s}$ where $\tilde{x}$ are waveform frames. We provide additional results with a Gaussian output distribution in appendix~\ref{app: additional likelihood results appendix}. 

We configure WaveNet as in the original paper using ten layers per block and five blocks. We use $D_c=96$ channels. 
We also train an LSTM model \citep{hochreiter_long_1997} which has fully connected encoders and decoders, as the VRNN and SRNN, but a deterministic recurrent cell and much fewer parameters than WaveNet. We report on LSTM models with $D_d=256$ hidden units. 

The configuration of the VRNN and SRNN models is similar to the LSTM. For both models we set the latent variable equal in size to the hidden units, $D_z=256$. At stack size $s=1$, the models are computationally demanding and hence we train them on randomly sampled segments from each training example and only on TIMIT.

The STCN is used in the ``dense" configuration of the original work \cite{aksan_stcn_2019}. It uses 256 convolutional channels and $L=5$ latent variables of dimensions $16, 32, 64, 128, 256$ from the top down. We also run a one-layered ablation with the same architecture but only one latent variable of dimension $256$ at the top. 
The CW-VAE is configured similar to the VRNN and SRNN models and with $c=8$. 
We run the CW-VAE with $L=1$ and $2$ layers of latent variables. 
The number of convolutional channels and is set equal to $D_z$ which is set to 96.

\paragraph{Baselines} We supply three elementary baselines that form approximate upper and lower bounds on the likelihood for arbitrary $s$. Specifically, we evaluate an uninformed per-frame discrete uniform distribution and a two-component DMoL fitted to the training set to benchmark worst case performance. 
We also report the likelihood achieved by the lossless compression algorithm, FLAC \citep{coalson_free_2019} which establishes a notion of good performance, although not a strict best case. We report FLAC on linear PCM since it was designed for this encoding.

\subsection{Likelihood results}
\begin{table}[t]
    \centering
    \begin{tabular}[t]{lll|r}
        \toprule
        $s$ & \bf Model         & \bf Configuration           & \bf $\mathcal{L}$ [bpf] \\
        \midrule
        1 & Uniform             & Uninformed                  & 16.00 \\
        1 & DMoL                & Optimal                     & 15.60 \\   %
        - & FLAC                & Linear PCM                  & \textbf{8.582} \\ %
        \midrule
        1 & WaveNet             & $D_c=96$                    & \textbf{10.88} \\  %
        1 & LSTM                & $D_d=256$                   & 10.97 \\  %
        1 & VRNN                & $D_z=256$                   & $\leq$11.09 \\  %
        1 & SRNN                & $D_z=256$                   & $\leq$11.19 \\  %
        1 & STCN-dense          & $D_z=256,L=5$               & $\leq$11.77 \\  %
        \midrule
        64 & WaveNet            & $D_c=96$                    & 13.30 \\  %
        64 & LSTM               & $D_d=256$                   & 13.34 \\  %
        64 & VRNN               & $D_z=256$                   & $\leq$12.54 \\  %
        64 & SRNN               & $D_z=256$                   & $\leq$12.42 \\  %
        64 & CW-VAE             & $D_z=96,L=1$                & $\leq$12.44 \\
        64 & CW-VAE             & $D_z=96,L=2$                & $\leq$12.17 \\
        64 & STCN-dense         & $D_z=256,L=1$               & $\leq$12.32 \\  %
        64 & STCN-dense         & $D_z=256,L=5$               & $\leq$\textbf{11.78} \\
        \midrule
        256 & WaveNet           & $D_c=96$                    & 14.11 \\  %
        256 & LSTM              & $D_d=256$                   & 14.20 \\  %
        256 & VRNN              & $D_z=256$                   & $\leq$13.27 \\  %
        256 & SRNN              & $D_z=256$                   & $\leq$13.14 \\  %
        256 & CW-VAE            & $D_z=96,L=1$                & $\leq$13.11 \\
        256 & CW-VAE            & $D_z=96,L=2$                & $\leq$12.97 \\
        256 & STCN-dense        & $D_z=256,L=1$               & $\leq$13.07 \\  %
        256 & STCN-dense        & $D_z=256,L=5$               & $\leq$\textbf{12.52} \\
        \bottomrule
    \end{tabular}
    \caption{
    Model likelihoods $\mathcal{L}$ on TIMIT represented as a 16bit $\mu$-law encoded PCM for different stochastic latent variable models compared to deterministic autoregressive baselines. 
    For the CW-VAE, $s$ refers to $s_1$ and the multi-layered models have $c=8$.
    Likelihoods are given in units of bits per frame (bpf).
    }
    \label{tab: likelihoods timit}
    \vspace{-0mm}
\end{table}

\begin{table*}[t]
    \centering
    \begin{tabular}{lll|rrrr}  %
        \toprule
        $s$ & \bf Model                   & \bf Configuration       & \multicolumn{4}{c}{\bf Likelihood $\mathcal{L}$ [bpf]} \\
         & & & dev-clean & dev-other & test-clean & test-other\\
         & & & 10h/100h & 10h/100h & 10h/100h & 10h/100h\\
        \midrule
        1 & Uniform             & Uninformed     & \multicolumn{1}{c}{16.00} & \multicolumn{1}{c}{16.00} & \multicolumn{1}{c}{16.00} & \multicolumn{1}{c}{16.00} \\
        1 & DMoL                & Optimal        & \multicolumn{1}{c}{15.66} & \multicolumn{1}{c}{15.70} & \multicolumn{1}{c}{15.62} & \multicolumn{1}{c}{15.71}\\   %
        - & FLAC                & Linear PCM     & \multicolumn{1}{c}{\bf 9.390} & \multicolumn{1}{c}{\bf 9.292} & \multicolumn{1}{c}{\bf 9.700} & \multicolumn{1}{c}{\bf 9.272} \\
        \midrule
        1 & Wavenet             & $D_c=96$       & \bf 10.96/10.89 & \bf 10.85/10.76 & \bf 11.12/11.01 & \bf 11.05/10.85 \\
        1 & LSTM                & $D_d=256$      & 11.21/11.17 & 11.10/11.06 & 11.35/11.29 & 11.28/11.23 \\  %
        \midrule
        64 & Wavenet            & $D_c=96$       & 13.61/13.24 & 13.58/13.21 & 13.61/13.22 & 13.60/13.21 \\  %
        64 & LSTM               & $D_d=256$      & 13.56/13.25 & 13.52/13.24 & 13.55/13.23 & 13.56/13.25 \\  %
        64 & CW-VAE             & $D_z=96,L=1$   & $\leq$12.32/12.24 & 12.32/12.23 & 12.43/12.33 & 12.43/12.33 \\
        64 & CW-VAE             & $D_z=96,L=2$   & $\leq$12.30/12.22 & 12.30/12.21 & 12.40/12.31 & 12.39/12.32 \\
        64 & STCN-dense         & $D_z=256,L=5$  & $\leq$\bf 11.98/11.47 & \bf 11.98/11.46 & \bf 12.08/11.58 & \bf 12.09/11.60 \\  %
        \bottomrule
    \end{tabular}
    \caption{
    Model likelihoods $\mathcal{L}$ on LibriSpeech test sets represented as $\SI{16}{bit}$ $\mu$-law encoded PCM.
    For the CW-VAE, $s$ refers to $s_1$ and the two-layered models have $s_2=8s_1$.
    The models are trained on either the $\SI{10}{h}$ LibriLight subset or the $\SI{100}{h}$ LibriSpeech \textrm{train-clean-100} subset as indicated.
    Likelihoods are given in units of bits per frame (bpf).
    }
    \label{tab: likelihoods librispeech}
    \vspace{-0mm}
\end{table*}

\begin{table}[t]
    \centering
    \begin{tabular}{cll|c}
        \toprule
        \multicolumn{3}{c}{\bf ASR configuration} & \multicolumn{1}{c}{\textbf{Result}} \\
Data   &  Model           & Input         &  PER [\%]  \\
        \midrule
3.7h   &  Spectrogram     & Mel           &   24.1   \\  %
3.7h   &  WaveNet         & $\vh^{(15)}$  &   27.7   \\  %
3.7h   &  LSTM            & $\vh$         &   23.0   \\  %
3.7h   &  VRNN            & $\vz$         &   23.2   \\  %
3.7h   &  SRNN            & $\vz$         &   26.0   \\  %
3.7h   &  CW-VAE          & $\vz^{(1)}$   &   36.4   \\  %
3.7h   &  STCN            & $\vz^{(2)}$   &   \textbf{21.9}   \\  %
\midrule
1.0h   &  Spectrogram     & Mel           &   30.8   \\  %
1.0h   &  WaveNet         & $\vh^{(15)}$  &   34.7   \\  %
1.0h   &  LSTM            & $\vh$         &   30.1   \\  %
1.0h   &  VRNN            & $\vz$         &   30.4   \\  %
1.0h   &  SRNN            & $\vz$         &   31.7   \\  %
1.0h   &  CW-VAE          & $\vz^{(1)}$   &   40.0   \\  %
1.0h   &  STCN            & $\vz^{(2)}$   &   \textbf{26.7}   \\  %
\midrule
10m    &  Spectrogram     & Mel           &   47.1   \\  %
10m    &  WaveNet         & $\vh^{(15)}$  &   52.8   \\  %
10m    &  LSTM            & $\vh$         &   46.1   \\  %
10m    &  VRNN            & $\vz$         &   44.6   \\  %
10m    &  SRNN            & $\vz$         &   47.3   \\  %
10m    &  CW-VAE          & $\vz^{(1)}$   &   54.9   \\  %
10m    &  STCN            & $\vz^{(2)}$   &   \textbf{42.7}   \\  %
\bottomrule
    \end{tabular}
    \caption{
        Evaluation of learned representations via phoneme recognition on TIMIT. 
        The ASR model is a three-layered bidirectional LSTM trained with CTC \cite{graves_connectionist_2006}. 
        The experiment is similar to that of \citet{hsu_unsupervised_2017} but we focus on the effect of the amount of labeled data and evaluate many more models. 
        The specific representations used is indicated in the input column.
    }
    \label{tab: phoneme recognition (PER)}
    \vspace{-0mm}
\end{table}

\begin{figure*}[t]
    \centering
    \includegraphics[width=0.555\textwidth]{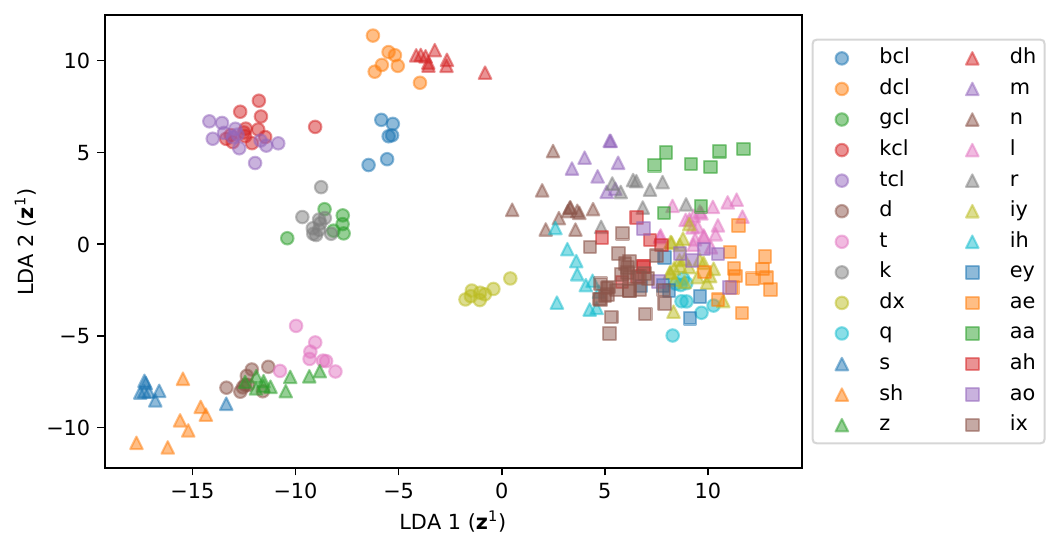}
    \hfill
    \includegraphics[width=0.425\textwidth]{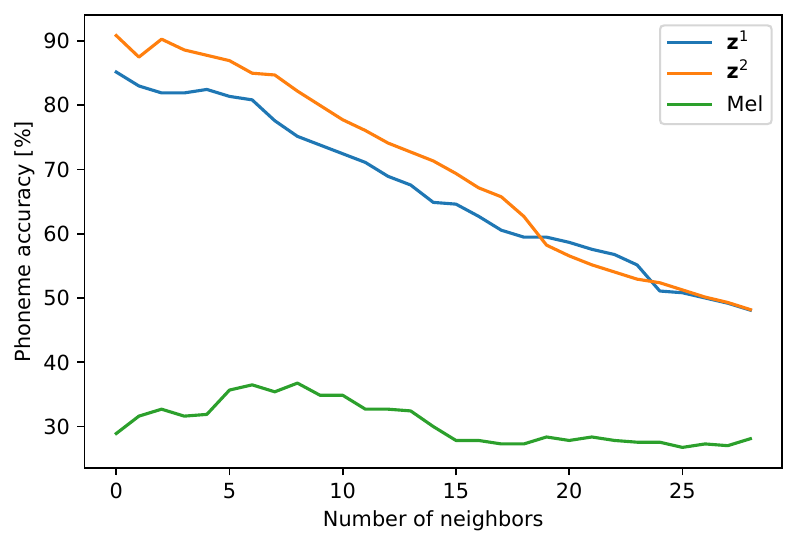}
    \caption{
    (left) Clustering of phonemes in a 2D Linear Discriminant Analysis (LDA) subspace of a CW-VAE latent space ($\vz^{(1)}$).
    (right) Leave-one-out phoneme classification accuracy for a KNN classifier at different $K$ in a 5D LDA subspace of a CW-VAE latent space.
    }
    \label{fig: latent space phoneme and knn}
    \vspace{-0mm}
\end{figure*}

\paragraph{TIMIT}
For temporal resolutions of $s=1$, the deterministic autoregressive models yield the best likelihoods with WaveNet achieving $\SI{10.88}{bpf}$ on TIMIT as seen in \tabref{tab: likelihoods timit}. 
Somewhat surprisingly, the LSTM baseline almost matches WaveNet with a likelihood of $\SI{11.11}{bpf}$ at $s=1$. However, due to being autoregressive in training, the LSTM trains considerably slower than the parallel convolutional WaveNet; something not conveyed by \tabref{tab: likelihoods timit}. 
Notably, the VRNN and SRNN models achieve likelihoods close to that of WaveNet and the LSTM at around $\SI{11.09}{bpf}$. The STCN exhibited instability when trained at $s=1$ and tended to diverge.

At $s=64$, Wavenet and the LSTM yield significantly worse likelihoods than all LVMs separated by $\sim\SI{1}{bit}$.
The CW-VAE outperforms the VRNN and SRNN when configured with a hierarchy of latent variables. 
With a single layer of latent variables, the CW-VAE is inferior to both SRNN and VRNN but notably still better than the LSTM. 
These observations carry to $s=256$, where a multilayered CW-VAE outperforms the LSTM, VRNN and SRNN. 
The STCN yields the best results at both $s=64$ and $s=256$. 
For strides $s>1$, previous work has attributed the inferior performance of autoregressive models without latent variables, such as WaveNet and the LSTM, to the ability of LVMs to model intra-step correlations \citep{lai_re-examination_2019}. 

Decreasing the resolution $s$ improves the likelihood for all LVMs. However, the best performing models, STCN and CW-VAE are not yet scalable to this regime for reasons related to numerical instability and computational infeasibility. This seems to indicate that LVMs may be able to outperform autoregressive models at $s=1$ in the future.

\paragraph{LibriSpeech}
On LibriSpeech (\tabref{tab: likelihoods librispeech}), results are similar to TIMIT. The STCN achieves the best likelihood at $s=64$ and the CW-VAE surpasses WaveNet and the LSTM. 

\paragraph{Compression}
A connection can be made between the model likelihoods and the compression rates of audio compression algorithms. 
Lossy compression algorithms, such as MP3, exploit the dynamic range of human hearing to achieve 70-95\% reduction in bit rate \citep{brandenburg_mpeg-2_1998} while lossless compression algorithms, such as FLAC, achieve 50-70\% reduction \citep{coalson_free_2019} independent of audio content. 
Although both the autoregressive models and the LVMs are lossy, their objective minimizes the amount of incurred loss. 
The best likelihoods reported in tables~\ref{tab: likelihoods timit} and \ref{tab: likelihoods librispeech} correspond to about a 30\% reduction in bit rate which indicates that there are significant gains in likelihood to be made in speech modeling.

\section{Phoneme recognition}
Although the likelihood is a practical metric for model comparison and selection, a high likelihood does not guarantee that a model has learned useful representations \cite{huszar_is_2017}. 
For speech data, we would expect models to learn features related to phonetics which would make them useful for tasks such as automatic speech recognition (ASR). 
The Mel spectrogram is a well-established representation of audio designed for speech recognition and is predefined rather than learned. %
To assess the usefulness of the representations learned by the benchmarked models, we compare them to the highly useful Mel spectrogram on the task of phoneme recognition. 
Phonemes are fundamental units of speech that relate to how parts of words are pronounced rather than characters or words themselves (see also appendix~\ref{app: timit phoneme distributions}).

\paragraph{Quantitative}
We train an ASR model to recognize phonemes and compare its performance when using input representations obtained from different unsupervised models. 
For WaveNet and the LSTM, we use the hidden state as the representation. For all LVMs, we use the latent variable. For the hierarchical LVMs and WaveNet, we run the experiment using each possible representation and report only the best one here. 
We compare the learned representations to a log Mel spectrogram with 80 filterbanks, hop length 64 and window size 128. 
We also compared to using raw PCM in vectors of 64 elements standardized to $[-1,1]$ but found that the ASR did not reliably converge at all which highlights the importance of input representation. 
The ASR model is a three-layered bidirectional LSTM with 256 hidden units. It is trained with the connectionist temporal classification (CTC) loss \cite{graves_connectionist_2006} which lets it jointly learn to align and classify without using label alignments. 
We pre-train all unsupervised models at $s=64$ on the full TIMIT training dataset excluding the validation data (3.7h) as in \tabref{tab: likelihoods timit}. 
We then train the ASR model on all 3.7h as well as 1h and 10m subsets. We report results on the test set in terms of phoneme error rates (PER) in \tabref{tab: phoneme recognition (PER)}. 

As expected, Mel spectrogram performs well achieving 24.1\% PER using 3.7 hours of labeled data. However, the ASR trained on STCN representations outperforms the Mel spectrogram with a PER of 21.9\%. 
This indicates that unsupervised STCN representations are phonetically rich and potentially better suited for speech modeling than the engineered Mel spectrogram. 
When the amount of labeled data is reduced, LVM representations suffer slightly less than deterministic ones. WaveNet representations are interestingly outperformed by both the LSTM and all LVMs.

\paragraph{Qualitative}
We qualitatively asses the learned latent representations selectively for the CW-VAE. 
We infer the latent variables of all utterances by a single speaker from the TIMIT test set. We sample the latent sequence $100$ times to estimate the mean representation per time step. We then compute the average latent representation over the duration of each phoneme using aligned phoneme labels. This approximately marginalizes out variation during the phoneme. We use linear discriminant analysis (LDA) \cite{fisher_use_1936} to obtain a low-dimensional linear subspace of the latent space. 
We visualize the resulting representations colored according to test set phoneme classes in \figref{fig: latent space phoneme and knn}. We note that many phonemes are separable in the linear subspace and that related phonemes such as ``s" and ``sh" are close.

We also show the average accuracy of a leave-one-out $k$-nearest-neighbor (KNN) classifier on the single left-out latent representation reduced with a 5-dimensional LDA as a function of the number of neighbors. 
We compare accuracy to a Mel-spectrogram averaged over each phoneme duration and LDA reduced. The spectrogram is computed with hop length set to 64, equal to $s_1$ for the CW-VAE, window size 256 and 80 Mel bins.
We see that both latent spaces yield significantly better KNN accuracies than the Mel features.

\section{Conclusion}
In this paper, we developed a benchmark for speech modeling with stochastic latent variable models (LVMs). 
We compared LVMs and deterministic autoregressive models on equal footing and found that LVMs achieve inferior likelihood compared to deterministic WaveNet and LSTM baselines. Surprisingly, the LSTM almost matched the popular WaveNet model. 
We saw that hierarchical LVMs, such as STCN and CW-VAE, outperformed non-hierarchical versions of themselves in ablation experiments as well as non-hierarchical LVMs such as VRNN and SRNN. This matches recent observations in the image domain. 
We noted that the STCN with an autoregressive decoder outperforms the non-autoregressive CW-VAE, which we adapted to speech. 
Finally, we found that LVMs can learn latent representations that are useful for phoneme recognition and better than Mel spectrograms, which are tailored for the task, when identical models are trained on top of the representations.
While the best performing models are not yet scalable to the highest temporal resolution, these results indicate that they might improve upon deterministic models in the future.

\newpage
\subsubsection*{Acknowledgments}
This research was partially funded by the Innovation Fund Denmark via the Industrial PhD Programme (grant no.\@ 0153-00167B). JF and SH were funded in part by the Novo Nordisk Foundation (grant no.\@ NNF20OC0062606) via the Center for Basic Machine Learning Research in Life Science (MLLS, \hyperlink{https://www.mlls.dk}{https://www.mlls.dk}). JF was further funded by the Novo Nordisk Foundation (grant no.\@ NNF20OC0065611) and the Independent Research Fund Denmark (grant no.\@ 9131-00082B). SH was further funded by VILLUM FONDEN (15334) and the European Research Council (ERC) under the European Union’s Horizon 2020 research and innovation programme (grant agreement no. 757360).

\bibliography{bibliography.bib}
\bibliographystyle{icml2022}

\newpage
\appendix
\onecolumn
\include{appendix}

\end{document}

%% file: math_commands.tex
\usepackage{amsmath,amsfonts,bm}

\def\figref#1{figure~\ref{#1}}

\def\tabref#1{table~\ref{#1}}
\def\twotabref#1#2{tables \ref{#1} and \ref{#2}}

\def\eqref#1{equation~\ref{#1}}

\def\1{\bm{1}}

\providecommand{\pa}[1]{{\left(#1\right)}}
\renewcommand{\pa}[1]{{\left(#1\right)}}
\providecommand{\bra}[1]{{\left[#1\right]}}
\renewcommand{\bra}[1]{{\left[#1\right]}}
\providecommand{\cbra}[1]{{\left\{#1\right\}}}
\renewcommand{\cbra}[1]{{\left\{#1\right\}}}

\def\valpha{{\bm{\alpha}}}
\def\vbeta{{\bm{\beta}}}

\def\vtheta{{\bm{\theta}}}
\def\vphi{{\bm{\phi}}}

\def\va{{\bm{a}}}

\def\vd{{\bm{d}}}
\def\ve{{\bm{e}}}

\def\vh{{\bm{h}}}

\def\vs{{\bm{s}}}

\def\vx{{\bm{x}}}

\def\vz{{\bm{z}}}

\DeclareMathAlphabet{\mathsfit}{\encodingdefault}{\sfdefault}{m}{sl}
\SetMathAlphabet{\mathsfit}{bold}{\encodingdefault}{\sfdefault}{bx}{n}

%% file: bayesnet_extras.tex
\usetikzlibrary{shapes}
\usetikzlibrary{fit}
\usetikzlibrary{chains}
\usetikzlibrary{arrows}

\tikzstyle{latent} = [circle,fill=latent-color,draw=black,inner sep=1pt,
minimum size=30pt, font=\fontsize{10}{10}\selectfont, node distance=1]
\tikzstyle{obs} = [latent,fill=observed-color]
\tikzstyle{const} = [rectangle, inner sep=0pt, node distance=1]
\tikzstyle{factor} = [rectangle, fill=black,minimum size=5pt, inner
sep=0pt, node distance=0.4]
\tikzstyle{det} = [latent, fill=deterministic-color, diamond, minimum size=35pt]

\tikzstyle{plate} = [draw, rectangle, rounded corners, fit=#1]
\tikzstyle{wrap} = [inner sep=0pt, fit=#1]
\tikzstyle{gate} = [draw, rectangle, dashed, fit=#1]

\tikzstyle{caption} = [font=\footnotesize, node distance=0] %
\tikzstyle{plate caption} = [caption, node distance=0, inner sep=0pt,
below left=5pt and 0pt of #1.south east] %
\tikzstyle{factor caption} = [caption] %
\tikzstyle{every label} += [caption] %

\tikzset{>={triangle 45}}

\renewcommand{\edge}[3][]{ %
  \foreach \x in {#2} { %
    \foreach \y in {#3} { %
      \path (\x) edge [->,#1] (\y) ;%
    } ;
  } ;
}

\renewcommand{\plate}[4][]{ %
  \node[wrap=#3] (#2-wrap) {}; %
  \node[plate caption=#2-wrap] (#2-caption) {#4}; %
  \node[plate=(#2-wrap)(#2-caption), #1] (#2) {}; %
}

%% file: appendix.tex
\section{Reproducibility statement}\label{app: reproducibility statement}
The source code used for the work presented in this paper will be made available before the conference. 
This code provides all details, practical and otherwise, needed to reproduce the results in this paper including data preprocessing, model training, model likelihood and latent space evaluation.
The source code also includes scripts for downloading and preparing the LibriSpeech, LibriLight and TIMIT datasets. The LibriSpeech and LibriLight datasets are open source and can be downloaded with the preparation scripts. They are also available at \url{https://www.openslr.org/12} and \url{https://github.com/facebookresearch/libri-light}, respectively. The TIMIT dataset is commercial and must be purchased and downloaded from \url{https://catalog.ldc.upenn.edu/LDC93S1} before running the preparation script.

The stochastic latent variable models considered in this work do not provide an exact likelihood estimate nor an exact latent space representation. For the likelihood, they provide a stochastic lower bound and some variation in the reproduced likelihoods as well as latent representations must be expected between otherwise completely identical forward passes. This variance is fairly small in practice when averaging over large datasets such as those considered in this work. We seed our experiments to reduce the randomness to a minimum, but parts of the algorithms underlying the CUDA framework are stochastic for efficiency. To retain computational feasibility, we  do not run experiments with a deterministic CUDA backend.

\section{Ethics statement}\label{app: ethics statement}
The work presented here fundamentally deals with automated perception of speech and generation of speech. These applications of machine learning potentially raise a number of ethical concerns. For instance, the these models might see possibly adverse use in automated surveillance and generation of deep fakes. To counter some of these effects, this work has focused on openness by using publicly available datasets for model development and benchmarking. Additionally, the work will open source the source code used to create these results. 
Ensuring the net positive effect of the development of these technologies is and must continue to be an ongoing effort.

We do not associate any significant ethical concerns with the datasets used in this work. However, one might note that the TIMIT dataset has somewhat skewed distributions in terms of gender and race diversity. Specifically, the male to female ratio is about two to one while the vast majority of speakers are Caucasian. Such statistics might have an effect of some ethical concern on downstream applications derived from such a dataset as also highlighted in recent research \citep{koenecke_racial_2020}. In LibriSpeech, there is an approximately equal number of female and male speakers while the diversity in race is unknown to the authors.

\section{Datasets}\label{app: dataset details}
\paragraph{TIMIT} 
TIMIT \citep{garofolo_timit_1993} is a speech dataset which contains $\SI{16}{kHz}$ recordings of 630 speakers of eight major dialects of American English, each reading ten phonetically rich sentences. It amounts to 6300 total recordings splits approximately in 3.94 hours of audio for training and 1.43 hours of audio for testing. No speakers or sentences in the test set are in the training set. 
The full train and test subsets of TIMIT are as in previous work \citep{chung_recurrent_2015, fraccaro_sequential_2016, aksan_stcn_2019}. 
We randomly sample 5\% of the training set to use as a validation set. 
TIMIT includes temporally aligned annotations of phonemes and words as well as speaker metadata such as gender, height, age, race, education level and dialect region \citep{garofolo_timit_1993}.

\paragraph{LibriSpeech and LibriLight}
The LibriSpeech dataset \citep{panayotov_librispeech_2015} consists of readings of public domain audio books amounting to approximately $\SI{1000}{h}$ of audio. The data is derived from the LibriVox project.
LibriLight \citep{kahn_libri-light_2020} is a subset of LibriSpeech created as an automatic speech transcription (ASR) benchmark with limited or no supervision.
We specifically train on the $\SI{100}{h}$ \textrm{train-clean-100} subset of LibriSpeech and the $\SI{10}{h}$ subset of LibriLight. 
In all cases we evaluate on all the test splits \textrm{dev-clean}, \textrm{dev-other}, \textrm{test-clean}, \textrm{test-other}.

Both datasets represent the audio as $\SI{16}{bit}$ pulse code modulation (PCM) sampled at $\SI{16000}{Hz}$.

\section{Model architectures}\label{app: model architectures}
This section details model architectures. See appendix~\ref{app: additional graphical models} for graphical models and appendix~\ref{app: training details} for training details. 

\paragraph{WaveNet}
We implement WaveNet as described in the original work \citep{oord_wavenet_2016} but use a discretized mixture of logistics as the output distribution as also done in other work \citep{oord_parallel_2018}. Our WaveNet is not conditioned on any signal other than the raw waveform. The model applies the causal convolution directly to the raw waveform frames (i.e. one input channel). An alternative option that we did not examine is to replace the initial convolution with an embedding lookup with a learnable vector for each waveform frame value.

\paragraph{LSTM}
The LSTM baseline uses an MLP encoder to embed the waveform subsegment $\vx_{t:t+s-1}$ to a feature vector before feeding it to the LSTM cell. The encoder is similar to the parameterization of $\phi_\text{vrnn}^\text{enc}$ for the VRNN described above. The LSTM cell produces the hidden state $\vd_t$ from $\vx_{t:t+s-1}$ and passes it to a decoder. Like the encoder, the decoder is parameterized like $\phi_\text{vrnn}^\text{dec}$ of the VRNN. It outputs the waveform predictions $\vx_{t+s:t+2s-1}$ from the hidden state $\vd_t$. The LSTM model uses a single vanilla unidirectional LSTM cell. 

\paragraph{VRNN}
We implement the VRNN as described in the original work \citep{chung_recurrent_2015} and verify that we can reproduce the original Gaussian likelihood TIMIT results. We replace the Gaussian output distribution with the DMoL. 

\paragraph{SRNN}
We implement the VRNN as described in the original work \citep{fraccaro_sequential_2016} and verify that we can reproduce the original Gaussian likelihood TIMIT results. We replace the Gaussian output distribution with the DMoL. 

\paragraph{CW-VAE} We implement the CW-VAE based on the original work \citep{saxena_clockwork_2021} but with some modifications also briefly described in section~\ref{sec: cwvae on speech}. We replace the encoder/decoder model architectures of the original work with architectures designed for waveform modeling. Specifically, the encoder and decoder are based on the Conv-TasNet \citep{luo_conv-tasnet_2019} and uses similar residual block structure. However, contrary to the Conv-TasNet, we require downsampling factors larger than two. In order to achieve this we use strides of two in the separable convolution of each block. With e.g. six blocks we hence get an overall stride of $2^6=64$. We can then add additional blocks with unit stride.
We also need to modify the residual connections that skip strided convolutions. Specifically, we replace the residual with a single convolution with stride equal to the stride used in the separable convolution. This convolution uses no nonlinearity and hence simply learns a local linear downsampling.

\paragraph{STCN} We implement the STCN as described in the original work \citep{aksan_stcn_2019} and verify that we can reproduce the original Gaussian likelihood TIMIT results. We replace the Gaussian output distribution with the DMoL. We use the best-performing version of the STCN reported in the original paper, namedly the ``STCN-dense" variant which conditions the observed variable on all five latent variables in the hierarchy. For the ablation experiment, we remove the bottom four latent variables. That is, we completely remove the corresponding four small densely connected networks that parameterize the prior and posterior distributions based on deterministic representations of the WaveNet encoder. We keep the top most prior and posterior networks and use them to parameterize a latent variable of $256$. This maintains the widest bottleneck of the model as well as almost all of the model's capacity.

\paragraph{ASR model}
The ASR model used for the phoneme recognition experiments is a three-layered bidirectional LSTM. We apply temporal dropout between the LSTM layers and also after the final layer. Temporal dropout works similar to regular dropout but samples the entries of the hidden state to mask only once and apply it to all timesteps, i.e. masking $\vh_{t}$ at vector index $i$ for all $t$ (and $i$). 
We mask by zeroing vector elements. We never mask the first timestep. 
We apply temporal dropout with masking probability of 0.3 for the 3.7h subset, 0.35 for the 1h subset and 0.4 for the 10m subset. 
The only difference in model architecture between the evaluation of different representations is the first affine transformation; from the dimensionality of the representation to the hidden state size of the LSTM. This gives rise to a very small difference in model capacity and parameter count which we find is negligible. We set the hidden unit size to 256.

\section{Training details}\label{app: training details}
\paragraph{Likelihood benchmark}
We implement all models and training scripts in PyTorch 1.9 \citep{paszke_automatic_2017}.
For both datasets we use the Adam optimizer \citep{kingma_adam_2015} with default parameters as given in PyTorch. We use learning rate $3\text{e}-4$ and no learning rate schedule.
We use PyTorch automatic mixed precision (AMP) to significantly reduce memory consumption. We did not observe any significant difference in final model performance compared to full ($\SI{32}{bit}$) precision.

We train stateful models (LSTM, VRNN, SRNN and CW-VAE) on the full sequence lengths padding batches with zeros when examples are not of equal length. We sample batches such that they consist of examples that are approximately the same length to minimize the amount of computation wasted on padding.

For $s=1$, we train stateless models (WaveNet, STCN) on random subsegments of the training examples and resample every epoch. This reduces memory requirements but does not bias the gradient. The subsequences are chosen to be of length 16000 which is larger than the receptive fields of the models and corresponds to one second of audio in TIMIT and LibriSpeech. 
For $s=64$ and $s=256$ we train the stateless models on the full example lengths similar to the stateful models since the receptive field is effectively $s$ times larger and the shorter sequence length reduces memory requirements.

In testing, we evaluate on the full sequences. Due to memory constrains, for LibriSpeech, we need to split the test examples into subsegments since the average sequence length in Librispeech is about 4 times longer than that of TIMIT. Hence, we do multiple forward passes per test example, one for each of several subsegments. We carry along the internal state for models that are autoregressive in training (LSTM, VRNN, SRNN, CW-VAE) and define segments to overlap according to model architecture.

\paragraph{Phoneme recognition}
The ASR experiment consists of two stages: 1) pre-training of the unsupervised model and 2) training of the ASR model. The pre-training is done as for the likelihood benchmark above. The ASR model is trained using the Adam optimizer \citep{kingma_adam_2015} with default parameters as given in PyTorch. We use learning rate $3\text{e}-4$ and no learning rate schedule. 

For the spectrogram, WaveNet and the LSTM, we extract the representation only once and train the ASR model on these. Since the models are deterministic and do not parameterize distributions, this is the only option. For the LVMs, we resample the latent representation of a training example at every epoch. This is the most principled approach as these models parameterize probability distributions. Furthermore, using a single sample would be subject to artificially high variance in the representations while it is not straightforward to establish a sound mean representation for sequential models.

\section{Converting the likelihood to units of bits per frame}\label{app: likelihood in bits per frame}
Here we briefly describe how to compute a likelihood in units of bits per frame (bpf). In the main text, we use $\log$ to mean $\log_e$, but here we will be explicit. In general, conversion from nats to bits (i.e., from $\log_e$ to $\log_2$) is achieved by $\log_2(x) = \log_e(x)  / \log_2(e) $. Remember that $\log_2 p(\vx_{1:T})$ generally factorizes as $\sum_t \log_2 p(x_t|\cdot)$. In sequence modeling, it is important to remember that each example $\vx^{i}$ must be weighted differently according the sequence length of that specific example. This is in contrast to computing bits per dimension in the image domain where images in a dataset are usually of the same dimensions. Thus, we compute the log-likelihood in bits per frame over the entire dataset as
\begin{equation}
    \mathcal{L}(\vx^{i}) = \frac{1}{\sum_i T_i} \sum_i \sum_t \log_2 p(\vx_t^{i}) \enspace ,
\end{equation}
where $i$ denotes the example index, $T_i$ is the length of example $\vx^i$ in waveform frames and $t$ is the time index. If a single timestep $\vx_t^i$ represents multiple waveform frames stacked with some stack size $s$, it is important to note that the sum over $t$ only has $T_i/s$ elements. 
For the LVMs, the term $\log_2 p(\vx_t^{i})$ is lower bounded by the ELBO in \eqref{eq: elbo}.

\section{Additional likelihood results} \label{app: additional likelihood results appendix}

\paragraph{TIMIT, $\mu$-law, DMoL} We provide additional results on TIMIT with audio represented as $\mu$-law encoded PCM in \tabref{tab: timit likelihoods dmol mu-law appendix}. Details are as presented in the main paper.

\paragraph{TIMIT, linear, DMoL}: We provide results on TIMIT with audio represented as linear PCM (raw PCM) in \tabref{tab: timit likelihoods dmol linear appendix}. Except for the encoding, details are as for $\mu$-law encoded TIMIT

\paragraph{TIMIT, linear, Gaussian} We also provide some results on TIMIT with the audio instead represented as linear PCM (linearly encoded) and using Gaussian output distributions as has been done previously in the literature \citep{chung_recurrent_2015, fraccaro_sequential_2016, lai_stochastic_2018, aksan_stcn_2019}. We use $s=200$ for comparability to the previous work. We provide the results in \tabref{tab: timit likelihoods gaussian appendix} and include likelihoods reported in the literature for reference. For our models, we use the same architectures as before but replace the discretized mixture of logistics with either a Gaussian distribution or a mixture of Gaussian distributions.

We constrain the variance of the Gaussians used with our models to be at least $\sigma^2_\text{min} = 0.01^2$ in order to avoid the variance going to zero, the likelihood going to infinity and optimization becoming unstable. The Gaussian standard deviation is clamped at minimum $0.001$ by \citet{aksan_stcn_2019}.

From \tabref{tab: timit likelihoods gaussian appendix} we note that the performance of the CW-VAE with Gaussian output distribution when modeling linear PCM (i.e. not $\mu$-law encoded) does not compare as favorably to the other baselines as it did with the discretized mixture of logistics distribution. We hypothesize that this has to do with using a Gaussian output distribution in latent variable models which, as has been reported elsewhere \citep{mattei_leveraging_2018}, leads to a likelihood function that is unbounded above and can grow arbitrarily high. We discuss this phenomenon in further detail in section~\ref{app: gaussian likelihood unboundedness discussion}. 

We specifically hypothesize that models that are autoregressive in the observed variable (VRNN, SRNN, Stochastic WaveNet, STCN) are well-equipped to utilize local smoothness to put very high density on the correct next value and that this in turn leads to a high degree of exploitation of the unboundedness of the likelihood. Not being autoregressive in the observed variable, the CW-VAE cannot exploit this local smoothness in the same way. Instead, the reconstruction is conditioned on a stochastic latent variable, $p(\vx_t|\vz^1_t)$, which introduces uncertainty and likely larger reconstruction variances.

\begin{table}[t!]
    \centering
    \begin{tabular}{ll|lr}
        \toprule
        $s$    & \bf Model           & \bf Configuration           & \bf $\mathcal{L}$ [bpf] \\
        \midrule
        1         & Uniform             & Uninformed            & 16.00 \\
        1         & DMoL                & Optimal               & 10.70 \\   %
        -         & FLAC                & Linear PCM            & 8.582 \\
        \midrule
        $1$       & Wavenet             & $D_C=96$              & \textbf{7.246} \\
        $1$       & LSTM                & $D_d=256, L=1$        & 7.295 \\
        $1$       & VRNN                & $D_z=256$             & $\leq$7.316 \\
        $1$       & SRNN                & $D_z=256$             & $\leq$7.501 \\
        $1$       & STCN                & $D_z=256,L=5$         & $\leq$9.970 \\
        \midrule
        $64$      & WaveNet             & $D_c=96$              & 8.402 \\
        $64$      & LSTM                & $D_d=256, L=1$        & 8.357 \\
        $64$      & VRNN                & $D_z=256$             & $\leq$8.103 \\
        $64$      & SRNN                & $D_z=256$             & $\leq$8.036 \\
        $64$      & CW-VAE              & $D_z=96, L=1$         & $\leq$7.989 \\
        $64$      & STCN                & $D_z=256,L=5$  & $\leq$\textbf{7.768} \\
        \midrule
        $256$      & WaveNet             & $D_c=96$              & 9.018 \\
        $256$      & LSTM                & $D_d=256, L=1$        & 8.959 \\  %
        $256$      & VRNN                & $D_z=256$             & $\leq$8.739 \\
        $256$      & SRNN                & $D_z=256$             & $\leq$8.674 \\
        $256$      & CW-VAE              & $D_z=96, L=1$         & $\leq$8.406 \\
        $256$      & STCN                & $D_z=256,L=5$  & $\leq$\textbf{8.196} \\
        \bottomrule
    \end{tabular}
    \caption{
    Model likelihoods on \textbf{TIMIT} represented as a \textbf{16 bit linear PCM}. The STCN converges to a poor local minimum and sometimes diverges when modeling linear PCM with $s=1$.
    }
    \label{tab: timit likelihoods dmol linear appendix}
\end{table}

\begin{table}[p]
    \centering
    \begin{tabular}{ll|lr}
        \toprule
        $s$    & \bf Model           & \bf Configuration           & \bf $\mathcal{L}$ [bpf] \\
        \midrule
        $1$       & Wavenet             & $D_C=16$              & 11.27 \\
        $1$       & Wavenet             & $D_C=24$              & 11.14 \\
        $1$       & Wavenet             & $D_C=32$              & 11.03 \\
        $1$       & Wavenet             & $D_C=96$              & 10.88 \\
        $1$       & Wavenet             & $D_C=128$             & 10.98 \\
        $1$       & Wavenet             & $D_C=160$             & 10.91 \\
        $1$       & LSTM                & $D_d=128, L=1$        & 11.40 \\
        $1$       & LSTM                & $D_d=256, L=1$        & 11.11 \\
        $1$       & VRNN                & $D_z=256$             & $\leq$11.09 \\
        $1$       & SRNN                & $D_z=256$             & $\leq$11.19 \\
        1 & STCN                & $D_z=256,L=5$               & $\leq$11.77 \\  %
        \midrule
        $4$       & LSTM                & $D_d=256, L=1$        & 11.65 \\
        \midrule
        $16$      & LSTM                & $D_d=256, L=1$        & 12.54 \\
        $16$      & LSTM                & $D_d=256, L=2$        & 12.54 \\
        $16$      & LSTM                & $D_d=256, L=3$        & 12.44 \\
        \midrule
        $64$      & WaveNet             & $D_c=96$              & 13.30 \\
        $64$      & LSTM                & $D_d=96, L=1$         & 13.49 \\
        $64$      & LSTM                & $D_d=96, L=2$         & 13.46 \\
        $64$      & LSTM                & $D_d=96, L=3$         & 13.40 \\
        $64$      & LSTM                & $D_d=256, L=1$        & 13.27 \\
        $64$      & LSTM                & $D_d=256, L=2$        & 13.29 \\
        $64$      & LSTM                & $D_d=256, L=3$        & 13.31 \\
        $64$      & LSTM                & $D_d=512, L=1$        & 13.37 \\
        $64$      & LSTM                & $D_d=512, L=2$        & 13.37 \\
        $64$      & LSTM                & $D_d=512, L=3$        & 13.41 \\
        $64$      & VRNN                & $D_z=96$              & $\leq$12.93 \\
        $64$      & VRNN                & $D_z=256$             & $\leq$12.54 \\
        $64$      & SRNN                & $D_z=96$              & $\leq$12.87 \\
        $64$      & SRNN                & $D_z=256$             & $\leq$12.42 \\
        $64$      & CW-VAE              & $D_z=96, L=1$         & $\leq$12.44 \\
        $64$      & CW-VAE              & $D_z=96, L=2$         & $\leq$12.17 \\
        $64$      & CW-VAE              & $D_z=96, L=3$         & $\leq$12.15 \\
        $64$      & CW-VAE              & $D_z=256, L=2$        & $\leq$12.10 \\
        $64$ & STCN               & $D_z=256,L=1$               & $\leq$12.32 \\  %
        $64$ & STCN               & $D_z=256,L=5$               & $\leq$11.78 \\
        \midrule
        $256$     & WaveNet             & $D_c=96$              & 14.11 \\
        $256$     & LSTM                & $D_d=256, L=1$        & 14.20 \\
        $256$     & LSTM                & $D_d=256, L=2$        & 14.17 \\
        $256$     & LSTM                & $D_d=256, L=3$        & 14.26 \\
        $256$     & VRNN                & $D_z=96$              & $\leq$13.51 \\
        $256$     & VRNN                & $D_z=256$             & $\leq$13.27 \\
        $256$     & SRNN                & $D_z=96$              & $\leq$13.28 \\
        $256$     & SRNN                & $D_z=256$             & $\leq$13.14 \\
        $256$     & CW-VAE              & $D_z=96, L=1$         & $\leq$13.11 \\
        $256$     & CW-VAE              & $D_z=96, L=2$         & $\leq$12.97 \\
        $256$     & CW-VAE              & $D_z=96, L=3$         & $\leq$12.87 \\
        $256$     & STCN                & $D_z=256,L=1$         & $\leq$13.07 \\  %
        $256$     & STCN                & $D_z=256,L=5$         & $\leq$12.52 \\
        \bottomrule
    \end{tabular}
    \caption{
    Model likelihoods on \textbf{TIMIT} represented as a \textbf{16 bit $\boldsymbol{\mu}$-law encoded PCM}.
    }
    \label{tab: timit likelihoods dmol mu-law appendix}
\end{table}

\begin{table}[t!]
    \centering
    \begin{tabular}{ll|lr}
        \toprule
        $s$ & \bf Model           & \bf Configuration           & \bf $\mathcal{L}$ [nats] \\
        \midrule
        1 & WaveNet                                                               & Normal & 119656 \\
        1 & WaveNet                                                               & GMM-2  & 120699 \\
        1 & WaveNet                                                               & GMM-20 & 121681 \\
        \midrule
        200 & WaveNet {\scriptsize \citep{aksan_stcn_2019}}                        & GMM-20    & 30188 \\
        200 & WaveNet {\scriptsize \citep{aksan_stcn_2019}}                        & Normal & -7443 \\
        200 & Stochastic WaveNet$^*$ {\scriptsize \citep{lai_stochastic_2018}}     & Normal & $\geq$72463\\
        200 & VRNN {\scriptsize \citep{chung_recurrent_2015}}                      & Normal & $\approx$28982\\
        200 & SRNN {\scriptsize \citep{fraccaro_sequential_2016}}                  & Normal & $\geq$60550 \\
        200 & STCN {\scriptsize \citep{aksan_stcn_2019}}                           & GMM-20 & $\geq$69195\\
        200 & STCN {\scriptsize \citep{aksan_stcn_2019}}                           & Normal & $\geq$64913\\
        200 & STCN-dense {\scriptsize \citep{aksan_stcn_2019}}                     & GMM-20 & $\geq$71386\\
        200 & STCN-dense {\scriptsize \citep{aksan_stcn_2019}}                     & Normal & $\geq$70294\\
        200 & STCN-dense-large {\scriptsize \citep{aksan_stcn_2019}}               & GMM-20 & $\geq$77438\\
        200 & CW-VAE$^*$                                         & $L=1, D_z=96$, Normal & $\geq$41629 \\  %
        \bottomrule
    \end{tabular}
    \caption{
    Model likelihoods on \textbf{TIMIT} represented as \textbf{globally normalized 16 bit linear PCM}. Contrary to the other likelihoods reported in this paper, here they are given in units of nats and obtained by summing the likelihood over time and over all examples in the dataset and dividing by the total number of examples. In the table, Normal refers to using a Gaussian likelihood and GMM refers to using a  Gaussian Mixture Model likelihood with 20 components. Models with asterisks $^*$ are our implementations while remaining results are as reported in the referenced work.
    }
    \label{tab: timit likelihoods gaussian appendix}
\end{table}

\section{Additional discussion on Gaussian likelihoods in LVMs} \label{app: gaussian likelihood unboundedness discussion}
As noted in section~\ref{app: additional likelihood results appendix}, we constrain the variance of the output distribution of our models to be $\sigma^2_\text{min} = 0.01^2$ for the additional results on TIMIT with Gaussian outputs. This limits the maximum value attainable by the prediction/reconstruction density of a single waveform frame $x_t$. 

Specifically, we can see that since
\begin{equation}
    \log p(x_t|\cdot) = \log \mathcal{N}\pa{x_t; \mu_t, \max\cbra{\sigma_\text{min}^2, \sigma_t^2}}\enspace \label{eq: gaussian ll with minimum variance} ,
\end{equation}
the best prediction/reconstruction density is achieved when $\sigma^2 \leq \sigma_\text{min}^2$ and $\mu=x_t$. 
Here $\cdot$ indicates any variables we might condition on such as the previous input frame $x_{t-1}$ or some latent variables.
We can evaluate this best case scenario for $\sigma_\text{min}^2 = 0.01^2$,
\begin{align}
    \log \mathcal{N}\pa{x_t; x_t, \sigma_\text{min}^2} &= -\frac{1}{2}\log 2\pi - \frac{1}{2}\log \sigma_\text{min}^2 - \frac{1}{2\sigma_\text{min}^2} (x_t - x_t) \nonumber \\
    &= -\frac{1}{2}\log 2\pi - \frac{1}{2}\log 0.01^2 \nonumber \\
    &= 3.686 \enspace . 
\end{align}
Hence, with perfect prediction/reconstruction and the minimal variance ($0.01^2$), a waveform frame contributes to the likelihood with $\SI{3.686}{nats}$. With an average test set example length of $\SI{49367.3}{frames}$ frames this leads to a best-case likelihood of 181967. We provide a list of maximally attainable Gaussian likelihoods on TIMIT for different minimal variances in \tabref{tab:timit best gaussian ll}. One can note that the maximal likelihood at $\sigma_\text{min}^2=0.1^2$ is lower than the likelihoods achieved by some models in \tabref{tab: timit likelihoods gaussian appendix}. This indicates that the models learn to use very small variances in order to increase the likelihood. Empirically, standard deviations smaller than approximately $0.001$ can result in numerical instability.

\begin{table}[t!]
    \centering
    \begin{tabular}{rrr}
    \toprule
    $\sigma_\text{min}$ & $\sigma_\text{min}^2$ & $\max\mathcal{L}$ \\
    \midrule
    $1$      & $1$           &    $-45367$ \\
    $0.5$    & $0.25$        &    $-11146$ \\
    $0.1$    & $0.01$        &     $68307$ \\
    $0.05$   & $0.0025$      &    $102525$ \\
    $0.01$   & $0.0001$      &    $181979$ \\
    $0.005$  & $0.000025$    &    $216198$ \\
    $0.001$  & $0.000001$    &    $295651$ \\
    \bottomrule
    \end{tabular}
    \caption{The highest possible Gaussian log-likelihoods (max $\mathcal{L}$) attainable on the TIMIT test set as computed by \eqref{eq: gaussian ll with minimum variance} with different values of the minimum variance $\sigma^2_\text{min}$.}
    \label{tab:timit best gaussian ll}
\end{table}

\section{Additional discussion on the choice of output distribution}\label{app: additional discussion on output distribution}
The DMoL uses a discretization of the continuous logistic distribution to define a mixture model over a discrete random variable. This allows it to parameterize multimodal distributions which can express ambiguity about the value of $\vx_t$.
The model can learn to maximize likelihood by assigning a bit of probability mass to multiple potential values of $\vx_t$.

While this is well-suited for autoregressive modeling, for which the distribution was developed, the potential multimodality poses a challenge for non-autoregressive latent variable models which independently sample multiple neighboring observations at the output.
In fact, if multiple neighboring outputs defined by the subsequence $\vx_{t_1:t_2}$ have multimodal $p(\vx_t|\cdot)$, we risk sampling a subsequence where each neighboring value expresses different potential realities, independently. 

Interestingly, most work on latent variable models with non-autoregressive output distributions seem to ignore this fact and simply employ the mixture distribution with 10 mixture components \citep{maaloe_biva_2019, vahdat_nvae_2020, child_very_2021}.
However, given the empirically good results of latent variable models for image generation, this seems to have posed only a minor problem in practice. We speculate that this is due to the high degree of similarity between neighbouring pixels in images. I.e. if the neighboring pixels are nuances of red, then, in all likelihood, so is the central pixel.

In the audio domain, however, neighbouring waveform frames can take wildly different values, especially at low sample rates. Furthermore, waveforms exhibit a natural symmetry between positive and negative amplitudes.
Hence, it seems plausible that multimodality may pose a larger problem in non-autoregressive speech generation by causing locally incoherent samples than it seems to do in image modelling.

\section{Additional graphical models}\label{app: additional graphical models}
In \figref{fig: cwvae cell state graph} we show the graphical model of the recurrent cell of the CW-VAE for a single time step. As noted in \citep{saxena_clockwork_2021}, this cell is very similar to the one of the Recurrent State Space Model (RSSM) \citep{hafner_learning_2019}.

In \figref{fig: cwvae three-layered graphical models (inference and generative)} we show the unrolled graphical models of a three-layered CW-VAE with $k_1=1$ and $c=2$ yielding $k_2=2$ and $k_3=4$. 
We show both the generative and inference models and highlight in blue the parameter sharing between the two models due to top-down inference.

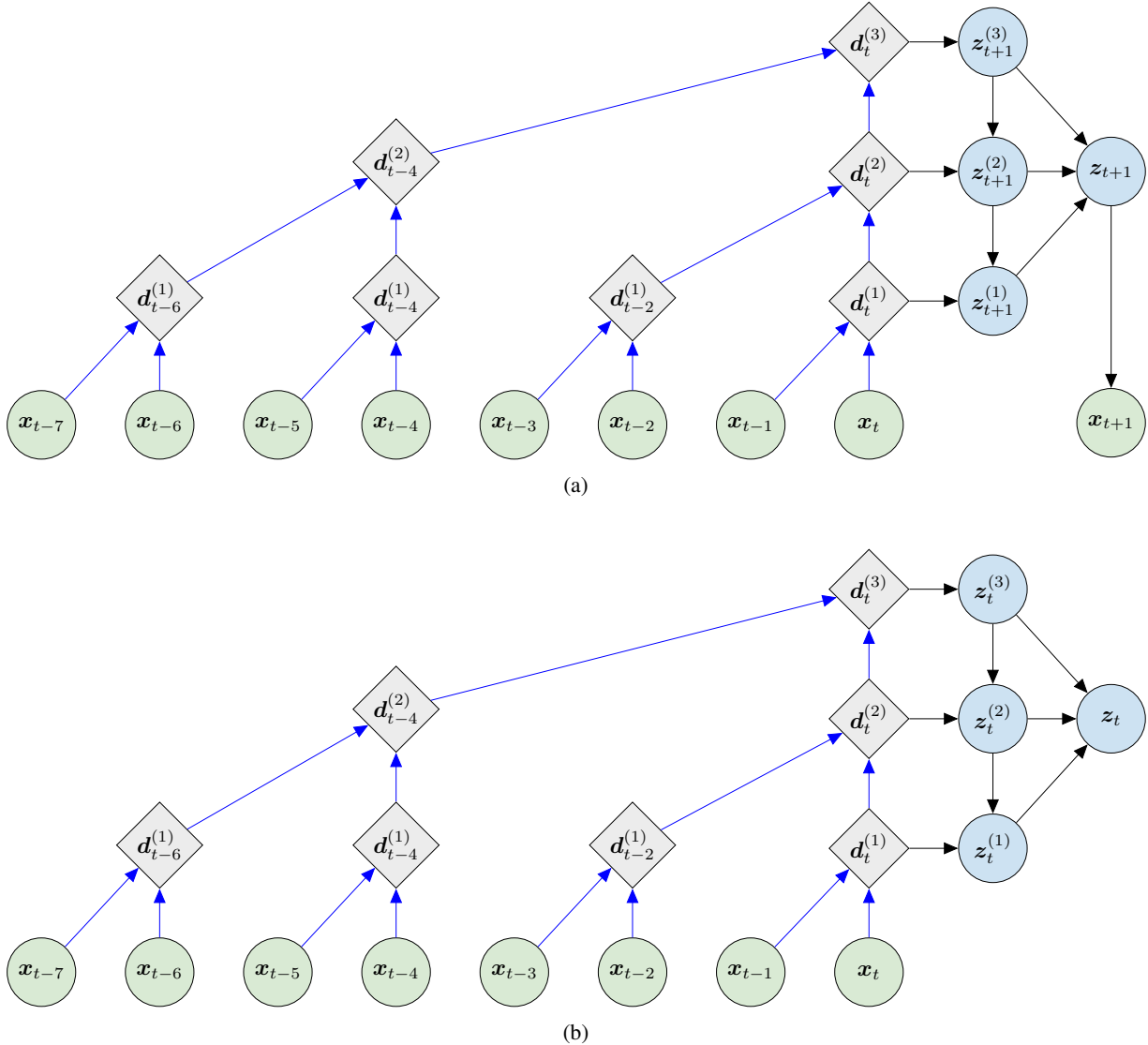
\begin{figure}[t!]
    \begin{subfigure}[b]{\textwidth}
    \centering
    \def\col{blue}
    \resizebox{!}{6.5cm}{
    \begin{tikzpicture}
        \node[obs] (x_1) {$\vx_t$};%
        \node[obs,left=.75cm of x_1] (x_2) {$\vx_{t-1}$};
        \node[obs,left=.75cm of x_2] (x_3) {$\vx_{t-2}$};
        \node[obs,left=.75cm of x_3] (x_4) {$\vx_{t-3}$};
        \node[obs,left=.75cm of x_4] (x_5) {$\vx_{t-4}$};
        \node[obs,left=.75cm of x_5] (x_6) {$\vx_{t-5}$};
        \node[obs,left=.75cm of x_6] (x_7) {$\vx_{t-6}$};
        \node[obs,left=.75cm of x_7] (x_8) {$\vx_{t-7}$};

        \node[det,above=.75cm of x_1](d_1_1){$\vd_{t}^{(1)}$};
        \node[det,above=.75cm of x_3](d_1_3){$\vd_{t-2}^{(1)}$};
        \node[det,above=.75cm of x_5](d_1_5){$\vd_{t-4}^{(1)}$};
        \node[det,above=.75cm of x_7](d_1_7){$\vd_{t-6}^{(1)}$};

        \node[det,above=.75cm of d_1_1](d_2_1){$\vd_{t}^{(2)}$};
        \node[det,above=.75cm of d_1_5](d_2_5){$\vd_{t-4}^{(2)}$};

        \node[det,above=.75cm of d_2_1](d_3_1){$\vd_{t}^{(3)}$};

        \node[latent,right=.75cm of d_1_1](z_1){$\vz_{t+1}^{(1)}$};
        \node[latent,right=.75cm of d_2_1](z_2){$\vz_{t+1}^{(2)}$};
        \node[latent,right=.75cm of d_3_1](z_3){$\vz_{t+1}^{(3)}$};

        \node[latent,right=.75cm of z_2](z){$\vz_{t+1}$};

        \node[obs,below=2.8cm of z] (x_0) {$\vx_{t+1}$};

        \edge[\col]{x_1}{d_1_1};
        \edge[\col]{x_2}{d_1_1};
        \edge[\col]{x_3}{d_1_3};
        \edge[\col]{x_4}{d_1_3};
        \edge[\col]{x_5}{d_1_5};
        \edge[\col]{x_6}{d_1_5};
        \edge[\col]{x_7}{d_1_7};
        \edge[\col]{x_8}{d_1_7};

        \edge[\col]{d_1_1}{d_2_1};
        \edge[\col]{d_1_3}{d_2_1};
        \edge[\col]{d_1_5}{d_2_5};
        \edge[\col]{d_1_7}{d_2_5};

        \edge[\col]{d_2_1}{d_3_1};
        \edge[\col]{d_2_5}{d_3_1};

        \edge[]{d_1_1}{z_1};
        \edge[]{d_2_1}{z_2};
        \edge[]{d_3_1}{z_3};

        \edge[]{z_3}{z_2}
        \edge[]{z_2}{z_1}

        \edge[]{z_1}{z}
        \edge[]{z_2}{z}
        \edge[]{z_3}{z}
        
        \edge[]{z}{x_0}

    \end{tikzpicture}
    }
    \caption{}
    \label{fig: stcn graphical model generative}
    \end{subfigure}
    \\
    \vspace{0.25cm}
    \\
    \begin{subfigure}[b]{\textwidth}
    \centering
    \def\col{blue}
    \resizebox{!}{6.5cm}{
    \begin{tikzpicture}
        \node[obs] (x_1) {$\vx_t$};%
        \node[obs,left=.75cm of x_1] (x_2) {$\vx_{t-1}$};
        \node[obs,left=.75cm of x_2] (x_3) {$\vx_{t-2}$};
        \node[obs,left=.75cm of x_3] (x_4) {$\vx_{t-3}$};
        \node[obs,left=.75cm of x_4] (x_5) {$\vx_{t-4}$};
        \node[obs,left=.75cm of x_5] (x_6) {$\vx_{t-5}$};
        \node[obs,left=.75cm of x_6] (x_7) {$\vx_{t-6}$};
        \node[obs,left=.75cm of x_7] (x_8) {$\vx_{t-7}$};

        \node[det,above=.75cm of x_1](d_1_1){$\vd_{t}^{(1)}$};
        \node[det,above=.75cm of x_3](d_1_3){$\vd_{t-2}^{(1)}$};
        \node[det,above=.75cm of x_5](d_1_5){$\vd_{t-4}^{(1)}$};
        \node[det,above=.75cm of x_7](d_1_7){$\vd_{t-6}^{(1)}$};
        
        \node[det,above=.75cm of d_1_1](d_2_1){$\vd_{t}^{(2)}$};
        \node[det,above=.75cm of d_1_5](d_2_5){$\vd_{t-4}^{(2)}$};
        
        \node[det,above=.75cm of d_2_1](d_3_1){$\vd_{t}^{(3)}$};
        
        \node[latent,right=.75cm of d_1_1](z_1){$\vz_{t}^{(1)}$};
        \node[latent,right=.75cm of d_2_1](z_2){$\vz_{t}^{(2)}$};
        \node[latent,right=.75cm of d_3_1](z_3){$\vz_{t}^{(3)}$};
        
        \node[latent,right=.75cm of z_2](z){$\vz_{t}$};
        
        \edge[\col]{x_1}{d_1_1};
        \edge[\col]{x_2}{d_1_1};
        \edge[\col]{x_3}{d_1_3};
        \edge[\col]{x_4}{d_1_3};
        \edge[\col]{x_5}{d_1_5};
        \edge[\col]{x_6}{d_1_5};
        \edge[\col]{x_7}{d_1_7};
        \edge[\col]{x_8}{d_1_7};

        \edge[\col]{d_1_1}{d_2_1};
        \edge[\col]{d_1_3}{d_2_1};
        \edge[\col]{d_1_5}{d_2_5};
        \edge[\col]{d_1_7}{d_2_5};
        
        \edge[\col]{d_2_1}{d_3_1};
        \edge[\col]{d_2_5}{d_3_1};
        
        \edge[]{d_1_1}{z_1};
        \edge[]{d_2_1}{z_2};
        \edge[]{d_3_1}{z_3};
        
        \edge[]{z_3}{z_2}
        \edge[]{z_2}{z_1}
        
        \edge[]{z_1}{z}
        \edge[]{z_2}{z}
        \edge[]{z_3}{z}
        
    \end{tikzpicture}
    }
    \caption{}
    \label{fig: stcn graphical model inference}
    \end{subfigure}
\caption{
    STCN \citep{aksan_stcn_2019} generative model $p(\vx,\vz)$ in (\subref{fig: stcn graphical model generative}) and inference model $q(\vz|\vx)$ in (\subref{fig: stcn graphical model inference}) for a single time-step.
    The WaveNet autoregressive encoder is shared between generative and inference models. It is depicted here with only one stack of three layers in order to illustrate the dilated convolution with limited space. In practice, the model uses ten layers in each of five stacks/cycles resulting in a much larger receptive field. Importantly, the model parameterizes the five latent variables using the last deterministic representation $\vd^{(l)}$ from each stack, i.e. only every fifth $l$ starting from $l=5$ and ending at $l=25$.
    Note that the generative model uses the prior to transform the WaveNet hidden states $\vd_{t}^{(l)}$ into the latent variable $\vz_{t+1}^{(l)}$ one step ahead in time compared to the approximate posterior which infers $\vz_{t}^{(l)}$.
    Also note that $\vz_t$ is constructed by concatenating all $\vz_t^{(l)}$. The original paper explores setting $\vz_t$ equal to $\vz_t^{(1)}$. The best-performing STCN for speech, which also the one we implement, uses a WaveNet decoder to predict $\vx_{t+1}$ from a sequence of $\vz_t$ rather than a per-timestep transform.
    Blue arrows are shared between the inference and generative models.
}
\label{fig: stcn graphical models (inference and generative)}
\end{figure}

In \figref{fig: stcn graphical models (inference and generative)} we show the graphical models of the STCN \cite{aksan_stcn_2019} at a single timestep. The model has three layers and shares the parameters of the WaveNet encoder between the inference and generative models.

In \figref{fig: vrnn graphical models (inference and generative)} we illustrate the unrolled graphical models of the inference and generative models of the VRNN \citep{chung_recurrent_2015}. We include the deterministic variable $\vd_t$ in order to illustrate the difference to other latent variable models.

Likewise, in \figref{fig: srnn graphical models (inference and generative)} we illustrate the unrolled graphical models the SRNN \citep{fraccaro_sequential_2016}.

\begin{figure}[t!]
    \centering
    \def\col{blue}
    \begin{tikzpicture}
        \node[det] (h_t_l) {$\vd_t^l$};
        \node[latent, below right=1cm of h_t_l] (z_t_l) {$\vz_t^l$};
        \node[latent, right=1.5cm of h_t_l] (s_t_l) {$\vs_t^l$};

        \plate [inner sep=0.4cm] {cell} {(h_t_l)(z_t_l)(s_t_l)} {};
        
        \node[latent,left=0.75cm of h_t_l] (s_tm1_l) {$\vs_{t-1}^l$};
        \node[latent,above=0.75cm of h_t_l] (s_t_lp1) {$\vs_t^{l+1}$};%
        \node[latent,below=1cm of z_t_l] (e_t_l) {$\ve_t^l$};%
        
        \node[right=1.0cm of s_t_l] (s_t_l_copy1) {};%
        \node[below=2.5cm of s_t_l] (s_t_l_copy2) {};%
        
        \edge[blue] {s_tm1_l} {h_t_l}
        \edge[blue] {s_t_lp1} {h_t_l}
        
        \edge[] {h_t_l} {z_t_l}
        \edge[dashed] {e_t_l} {z_t_l}
        
        \edge[blue] {h_t_l} {s_t_l}
        \edge[blue] {z_t_l} {s_t_l}
        
        \edge[blue] {s_t_l} {s_t_l_copy1}
        \edge[blue] {s_t_l} {s_t_l_copy2}
    \end{tikzpicture}
\caption{CW-VAE cell state $\vs_t^l$ update. The cell state is given as $\vs_t^l=(\vz_t^l, \vd_t^l)$ where $\vd_t^l$ is the deterministic hidden state of a Gated Recurrent Unit \citep{cho_properties_2014}. The vector $\ve_t^l$ is computed from $\vx_t$ by the encoder network which outputs $L$ encodings, one for each latent variable, similar to that of a Ladder VAE \cite{sonderby_ladder_2016}. All blue arrows are shared between generation and inference. The dashed arrow is used only during inference. The solid arrow has unique transformations during inference and generation.}
\label{fig: cwvae cell state graph}
\end{figure}
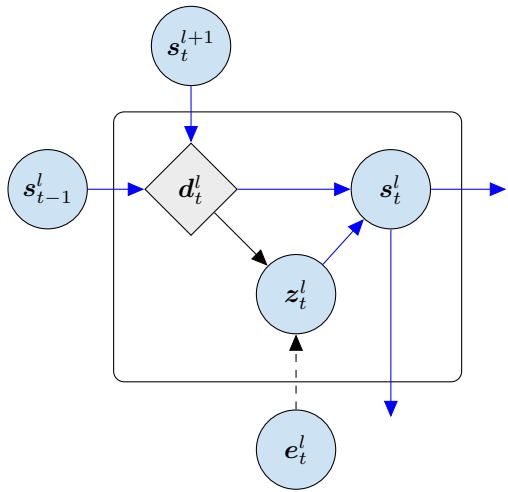

\begin{figure}[t!]
    \begin{subfigure}[b]{\textwidth}
    \flushright
    \def\col{blue}
    \resizebox{!}{6cm}{
    \begin{tikzpicture}
        \node[obs] (x_1) {$\vx_t$};%
        \node[obs,right=1.00cm of x_1] (x_2) {$\vx_{t+1}$};%
        \node[obs,right=1.00cm of x_2] (x_3) {$\vx_{t+2}$};%
        \node[obs,right=1.00cm of x_3] (x_4) {$\vx_{t+3}$};%
        \node[obs,right=1.00cm of x_4] (x_5) {$\vx_{t+4}$};%
        \node[obs,right=1.00cm of x_5] (x_6) {$\vx_{t+5}$};%
        \node[obs,right=1.00cm of x_6] (x_7) {$\vx_{t+6}$};%
        \node[obs,right=1.00cm of x_7] (x_8) {$\vx_{t+7}$};%
    
        \node[latent,above=1.00cm of x_1](s_1_1){$\vz^{(1)}_{t}$}; %
        \node[latent,above=1.00cm of x_2](s_1_2){$\vz^{(1)}_{t+1}$}; %
        \node[latent,above=1.00cm of x_3](s_1_3){$\vz^{(1)}_{t+2}$}; %
        \node[latent,above=1.00cm of x_4](s_1_4){$\vz^{(1)}_{t+3}$}; %
        \node[latent,above=1.00cm of x_5](s_1_5){$\vz^{(1)}_{t+4}$}; %
        \node[latent,above=1.00cm of x_6](s_1_6){$\vz^{(1)}_{t+5}$}; %
        \node[latent,above=1.00cm of x_7](s_1_7){$\vz^{(1)}_{t+6}$}; %
        \node[latent,above=1.00cm of x_8](s_1_8){$\vz^{(1)}_{t+7}$}; %
    
        \node[latent,above=1.00cm of s_1_1](s_2_1){$\vz^{(2)}_{t}$}; %
        \node[latent,above=1.00cm of s_1_3](s_2_3){$\vz^{(2)}_{t+2}$}; %
        \node[latent,above=1.00cm of s_1_5](s_2_5){$\vz^{(2)}_{t+4}$}; %
        \node[latent,above=1.00cm of s_1_7](s_2_7){$\vz^{(2)}_{t+6}$}; %
    
        \node[latent,above=1.00cm of s_2_1](s_3_1){$\vz^{(3)}_{t}$}; %
        \node[latent,above=1.00cm of s_2_5](s_3_5){$\vz^{(3)}_{t+4}$}; %
    
        \edge[]{s_1_1}{x_1};
        \edge[]{s_1_2}{x_2};
        \edge[]{s_1_3}{x_3};
        \edge[]{s_1_4}{x_4};
        \edge[]{s_1_5}{x_5};
        \edge[]{s_1_6}{x_6};
        \edge[]{s_1_7}{x_7};
        \edge[]{s_1_8}{x_8};
    
        \edge[blue]{s_1_1}{s_1_2};
        \edge[blue]{s_1_2}{s_1_3};
        \edge[blue]{s_1_3}{s_1_4};
        \edge[blue]{s_1_4}{s_1_5};
        \edge[blue]{s_1_5}{s_1_6};
        \edge[blue]{s_1_6}{s_1_7};
        \edge[blue]{s_1_7}{s_1_8};
        
        \edge[blue]{s_2_1}{s_2_3};
        \edge[blue]{s_2_3}{s_2_5};
        \edge[blue]{s_2_5}{s_2_7};
        
        \edge[blue]{s_3_1}{s_3_5};
        
        \edge[blue]{s_2_1}{s_1_1};
        \edge[blue]{s_2_3}{s_1_3};
        \edge[blue]{s_2_5}{s_1_5};
        \edge[blue]{s_2_7}{s_1_7};
    
        \edge[blue]{s_3_1}{s_2_1};
        \edge[blue]{s_3_5}{s_2_5};
        
        \edge[blue]{s_2_1}{s_1_2};
        \edge[blue]{s_2_3}{s_1_4};
        \edge[blue]{s_2_5}{s_1_6};
        \edge[blue]{s_2_7}{s_1_8};
        
        \edge[blue]{s_3_1}{s_2_3};
        \edge[blue]{s_3_5}{s_2_7};
    
    \end{tikzpicture}
    \hspace{2cm}
    }
    \caption{}
    \label{fig: cwvae three-layered graphical model generative}
    \end{subfigure}
    \\
    \vspace{0.25cm}
    \\
    \begin{subfigure}[b]{\textwidth}
    \flushright
    \def\col{blue}
    \resizebox{!}{6cm}{
    \begin{tikzpicture}
        \node[obs] (x_1) {$\vx_t$};%
        \node[obs,right=1.00cm of x_1] (x_2) {$\vx_{t+1}$};%
        \node[obs,right=1.00cm of x_2] (x_3) {$\vx_{t+2}$};%
        \node[obs,right=1.00cm of x_3] (x_4) {$\vx_{t+3}$};%
        \node[obs,right=1.00cm of x_4] (x_5) {$\vx_{t+4}$};%
        \node[obs,right=1.00cm of x_5] (x_6) {$\vx_{t+5}$};%
        \node[obs,right=1.00cm of x_6] (x_7) {$\vx_{t+6}$};%
        \node[obs,right=1.00cm of x_7] (x_8) {$\vx_{t+7}$};%
    
        \node[latent,above=1.00cm of x_1](s_1_1){$\vz^{(1)}_{t}$}; %
        \node[latent,above=1.00cm of x_2](s_1_2){$\vz^{(1)}_{t+1}$}; %
        \node[latent,above=1.00cm of x_3](s_1_3){$\vz^{(1)}_{t+2}$}; %
        \node[latent,above=1.00cm of x_4](s_1_4){$\vz^{(1)}_{t+3}$}; %
        \node[latent,above=1.00cm of x_5](s_1_5){$\vz^{(1)}_{t+4}$}; %
        \node[latent,above=1.00cm of x_6](s_1_6){$\vz^{(1)}_{t+5}$}; %
        \node[latent,above=1.00cm of x_7](s_1_7){$\vz^{(1)}_{t+6}$}; %
        \node[latent,above=1.00cm of x_8](s_1_8){$\vz^{(1)}_{t+7}$}; %
    
        \node[latent,above=1.00cm of s_1_1](s_2_1){$\vz^{(2)}_{t}$}; %
        \node[latent,above=1.00cm of s_1_3](s_2_3){$\vz^{(2)}_{t+2}$}; %
        \node[latent,above=1.00cm of s_1_5](s_2_5){$\vz^{(2)}_{t+4}$}; %
        \node[latent,above=1.00cm of s_1_7](s_2_7){$\vz^{(2)}_{t+6}$}; %
    
        \node[latent,above=1.00cm of s_2_1](s_3_1){$\vz^{(3)}_{t}$}; %
        \node[latent,above=1.00cm of s_2_5](s_3_5){$\vz^{(3)}_{t+4}$}; %
    
        \edge[]{x_1}{s_1_1};
        \edge[]{x_2}{s_1_2};
        \edge[]{x_3}{s_1_3};
        \edge[]{x_4}{s_1_4};
        \edge[]{x_5}{s_1_5};
        \edge[]{x_6}{s_1_6};
        \edge[]{x_7}{s_1_7};
        \edge[]{x_8}{s_1_8};
    
        \edge[bend left=40]{x_1}{s_2_1};
        \edge[bend left=40]{x_1}{s_3_1};
        \edge[]{x_2}{s_2_1};
        \edge[]{x_2}{s_3_1};
        \edge[bend left=40]{x_3}{s_2_3};
        \edge[]{x_3}{s_3_1};
        \edge[]{x_4}{s_2_3};
        \edge[]{x_4}{s_3_1};
        \edge[bend left=40]{x_5}{s_2_5};
        \edge[bend left=40]{x_5}{s_3_5};
        \edge[]{x_6}{s_2_5};
        \edge[]{x_6}{s_3_5};
        \edge[bend left=40]{x_7}{s_2_7};
        \edge[]{x_7}{s_3_5};
        \edge[]{x_8}{s_2_7};
        \edge[]{x_8}{s_3_5};
    
        \edge[\col]{s_1_1}{s_1_2};
        \edge[\col]{s_1_2}{s_1_3};
        \edge[\col]{s_1_3}{s_1_4};
        \edge[\col]{s_1_4}{s_1_5};
        \edge[\col]{s_1_5}{s_1_6};
        \edge[\col]{s_1_6}{s_1_7};
        \edge[\col]{s_1_7}{s_1_8};
        
        \edge[\col]{s_2_1}{s_2_3};
        \edge[\col]{s_2_3}{s_2_5};
        \edge[\col]{s_2_5}{s_2_7};
        
        \edge[\col]{s_3_1}{s_3_5};
        
        \edge[\col]{s_2_1}{s_1_1};
        \edge[\col]{s_2_3}{s_1_3};
        \edge[\col]{s_2_5}{s_1_5};
        \edge[\col]{s_2_7}{s_1_7};
    
        \edge[\col]{s_3_1}{s_2_1};
        \edge[\col]{s_3_5}{s_2_5};
        
        \edge[\col]{s_2_1}{s_1_2};
        \edge[\col]{s_2_3}{s_1_4};
        \edge[\col]{s_2_5}{s_1_6};
        \edge[\col]{s_2_7}{s_1_8};
        
        \edge[\col]{s_3_1}{s_2_3};
        \edge[\col]{s_3_5}{s_2_7};
    
    \end{tikzpicture}
    \hspace{2cm}
    }
    \caption{}
    \label{fig: cwvae three-layered graphical model inference}
    \end{subfigure}
\caption{
    CW-VAE \citep{saxena_clockwork_2021} generative model $p(\vx,\vz)$ in (\subref{fig: cwvae three-layered graphical model generative}) and inference model $q(\vz|\vx)$ in (\subref{fig: cwvae three-layered graphical model inference}) for a three-layered model with $k_1=1$ and $c=2$ giving $k_2=2$ and $k_3=4$ unrolled over eight steps in the observed variable. Blue arrows are (mostly) shared between the inference and generative models. See \figref{fig: cwvae cell state graph} for a detailed graphical model expanding on the latent nodes $\vz^l_t$ and parameter sharing.
}
\label{fig: cwvae three-layered graphical models (inference and generative)}
\end{figure}

\begin{figure}[t!]
    \centering
    \hfill
    \def\col{blue}
    \begin{subfigure}[b]{0.48\textwidth}
    \centering
    \begin{tikzpicture}
        \node[obs] (x_1) {$\vx_t$};%
        \node[obs,right=1.00cm of x_1] (x_2) {$\vx_{t+1}$};
        \node[obs,right=1.00cm of x_2] (x_3) {$\vx_{t+2}$};

        \node[latent,above=1.00cm of x_1](z_1_1){$\vz_{t}$};
        \node[latent,above=1.00cm of x_2](z_1_2){$\vz_{t+1}$};
        \node[latent,above=1.00cm of x_3](z_1_3){$\vz_{t+2}$};
        
        \node[det,above=1.00cm of z_1_1](d_1_1){$\vd_{t}$};
        \node[det,above=1.00cm of z_1_2](d_1_2){$\vd_{t+1}$};
        \node[det,above=1.00cm of z_1_3](d_1_3){$\vd_{t+2}$};

        \edge[]{z_1_1}{x_1};
        \edge[]{z_1_2}{x_2};
        \edge[]{z_1_3}{x_3};
        
        \edge[\col]{d_1_1}{z_1_1};
        \edge[\col]{d_1_2}{z_1_2};
        \edge[\col]{d_1_3}{z_1_3};
        
        \edge[bend left]{d_1_1}{x_1};
        \edge[bend left]{d_1_2}{x_2};
        \edge[bend left]{d_1_3}{x_3};
                
        \edge[\col]{d_1_1}{d_1_2};
        \edge[\col]{d_1_2}{d_1_3};
        
        \edge[\col]{x_1.35}{d_1_2};
        \edge[\col]{x_2.35}{d_1_3};
        
        \edge[\col]{z_1_1}{d_1_2};
        \edge[\col]{z_1_2}{d_1_3};

        \node[above=of d_1_2, yshift=-1.cm] (phi) {$p_\theta(\vz,\vx)$};
    \end{tikzpicture}
    \caption{}
    \label{fig: vrnn graphical model generative}
    \end{subfigure}
    \begin{subfigure}[b]{0.48\textwidth}
    \centering
    \begin{tikzpicture}
        \node[obs] (x_1) {$\vx_t$};%
        \node[obs,right=1.00cm of x_1] (x_2) {$\vx_{t+1}$};
        \node[obs,right=1.00cm of x_2] (x_3) {$\vx_{t+2}$};

        \node[latent,above=1.00cm of x_1](z_1_1){$\vz_{t}$};
        \node[latent,above=1.00cm of x_2](z_1_2){$\vz_{t+1}$};
        \node[latent,above=1.00cm of x_3](z_1_3){$\vz_{t+2}$};
            
        \node[det,above=1.00cm of z_1_1](d_1_1){$\vd_{t}$};
        \node[det,above=1.00cm of z_1_2](d_1_2){$\vd_{t+1}$};
        \node[det,above=1.00cm of z_1_3](d_1_3){$\vd_{t+2}$};

        \edge[]{x_1}{z_1_1};
        \edge[]{x_2}{z_1_2};
        \edge[]{x_3}{z_1_3};
        
        \edge[\col]{d_1_1}{z_1_1};
        \edge[\col]{d_1_2}{z_1_2};
        \edge[\col]{d_1_3}{z_1_3};
                
        \edge[\col]{d_1_1}{d_1_2};
        \edge[\col]{d_1_2}{d_1_3};
        
        \edge[\col]{x_1}{d_1_2};
        \edge[\col]{x_2}{d_1_3};
        
        \edge[\col]{z_1_1}{d_1_2};
        \edge[\col]{z_1_2}{d_1_3};

        \node[above=of d_1_2, yshift=-1.cm] (phi) {$q_\phi(\vz|\vx)$};
    \end{tikzpicture}
    \caption{}
    \label{fig: vrnn graphical model inference}
    \end{subfigure}
\caption{
    VRNN \citep{chung_recurrent_2015} generative model $p(\vx,\vz)$ in (\subref{fig: vrnn graphical model generative}) and inference model $q(\vz|\vx)$ in (\subref{fig: vrnn graphical model inference}) unrolled over three steps in the observed variable. 
    Blue arrows are shared between the inference and generative models.
}
\label{fig: vrnn graphical models (inference and generative)}
\end{figure}
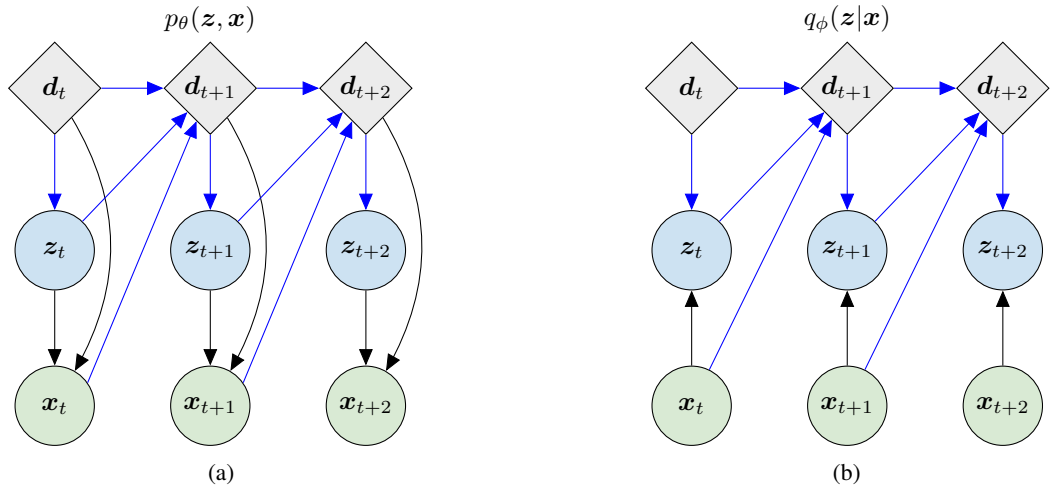

\begin{figure}[t!]
    \centering
    \hfill
    \def\col{blue}
    \begin{subfigure}[b]{0.48\textwidth}
    \centering
    \begin{tikzpicture}
        \node[obs] (x_1) {$\vx_t$};%
        \node[obs,right=1.00cm of x_1] (x_2) {$\vx_{t+1}$};
        \node[obs,right=1.00cm of x_2] (x_3) {$\vx_{t+2}$};

        \node[latent,above=1.00cm of x_1](z_1_1){$\vz_{t}$};
        \node[latent,above=1.00cm of x_2](z_1_2){$\vz_{t+1}$};
        \node[latent,above=1.00cm of x_3](z_1_3){$\vz_{t+2}$};
        
        \node[det,above=1.00cm of z_1_1](d_1_1){$\vd_{t}$};
        \node[det,above=1.00cm of z_1_2](d_1_2){$\vd_{t+1}$};
        \node[det,above=1.00cm of z_1_3](d_1_3){$\vd_{t+2}$};

        \edge[]{z_1_1}{x_1};
        \edge[]{z_1_2}{x_2};
        \edge[]{z_1_3}{x_3};
        
        \edge[]{d_1_1}{z_1_1};
        \edge[]{d_1_2}{z_1_2};
        \edge[]{d_1_3}{z_1_3};
        
        \edge[bend left=35]{d_1_1}{x_1};
        \edge[bend left=35]{d_1_2}{x_2};
        \edge[bend left=35]{d_1_3}{x_3};
                
        \edge[\col]{d_1_1}{d_1_2};
        \edge[\col]{d_1_2}{d_1_3};
        
        \edge[\col]{x_1.35}{d_1_2};
        \edge[\col]{x_2.35}{d_1_3};
        
        \edge[\col]{z_1_1}{z_1_2};
        \edge[\col]{z_1_2}{z_1_3};
        
        \node[above=of d_1_2, yshift=-1.cm] (phi) {$p_\theta(\vz,\vx)$};
    \end{tikzpicture}
    \caption{}
    \label{fig: srnn graphical model generative}
    \end{subfigure}
    \begin{subfigure}[b]{0.48\textwidth}
    \centering
    \begin{tikzpicture}
        \node[obs] (x_1) {$\vx_t$};%
        \node[obs,right=1.00cm of x_1] (x_2) {$\vx_{t+1}$};
        \node[obs,right=1.00cm of x_2] (x_3) {$\vx_{t+2}$};

        \node[latent,above=1.00cm of x_1](z_1_1){$\vz_{t}$};
        \node[latent,above=1.00cm of x_2](z_1_2){$\vz_{t+1}$};
        \node[latent,above=1.00cm of x_3](z_1_3){$\vz_{t+2}$};

        \node[det,above=1.00cm of z_1_1](a_1_1){$\va_{t}$};
        \node[det,above=1.00cm of z_1_2](a_1_2){$\va_{t+1}$};
        \node[det,above=1.00cm of z_1_3](a_1_3){$\va_{t+2}$};

        \node[det,above=1.00cm of a_1_1](d_1_1){$\vd_{t}$};
        \node[det,above=1.00cm of a_1_2](d_1_2){$\vd_{t+1}$};
        \node[det,above=1.00cm of a_1_3](d_1_3){$\vd_{t+2}$};

        \edge[bend left=40]{x_1}{a_1_1};
        \edge[bend left=40]{x_2}{a_1_2};
        \edge[bend left=40]{x_3}{a_1_3};
        
        \edge[\col]{x_1}{d_1_2};
        \edge[\col]{x_2}{d_1_3};
        
        \edge[]{d_1_1}{a_1_1};
        \edge[]{d_1_2}{a_1_2};
        \edge[]{d_1_3}{a_1_3};

        \edge[]{a_1_1}{z_1_1};
        \edge[]{a_1_2}{z_1_2};
        \edge[]{a_1_3}{z_1_3};

        \edge[\col]{d_1_1}{d_1_2};
        \edge[\col]{d_1_2}{d_1_3};

        \edge[]{a_1_2}{a_1_1};
        \edge[]{a_1_3}{a_1_2};

        \edge[\col]{z_1_1}{z_1_2};
        \edge[\col]{z_1_2}{z_1_3};

        \node[above=of d_1_2, yshift=-1.cm] (phi) {$q_\phi(\vz|\vx)$};
    \end{tikzpicture}
    \caption{}
    \label{fig: srnn graphical model inference}
    \end{subfigure}
\caption{
    SRNN \citep{fraccaro_sequential_2016} generative model $p(\vx,\vz)$ in (\subref{fig: srnn graphical model generative}) and inference model $q(\vz|\vx)$ in (\subref{fig: srnn graphical model inference}) unrolled over three steps in the observed variable. 
    Blue arrows are shared between the inference and generative models.
}
\label{fig: srnn graphical models (inference and generative)}
\end{figure}

\section{Additional latent evaluation}\label{app: additional latent space clustering}
\begin{figure*}[t!]
     \centering
     \includegraphics[width=0.477\textwidth]{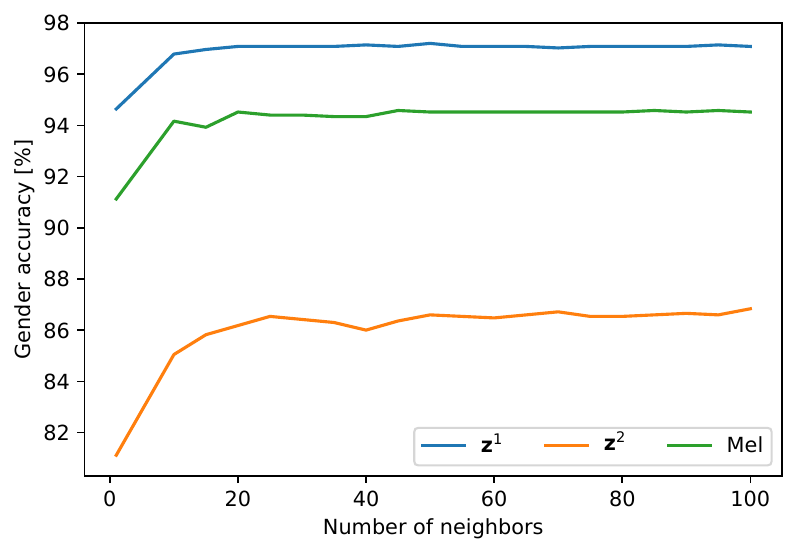}
     \hfill
     \includegraphics[width=0.483\textwidth]{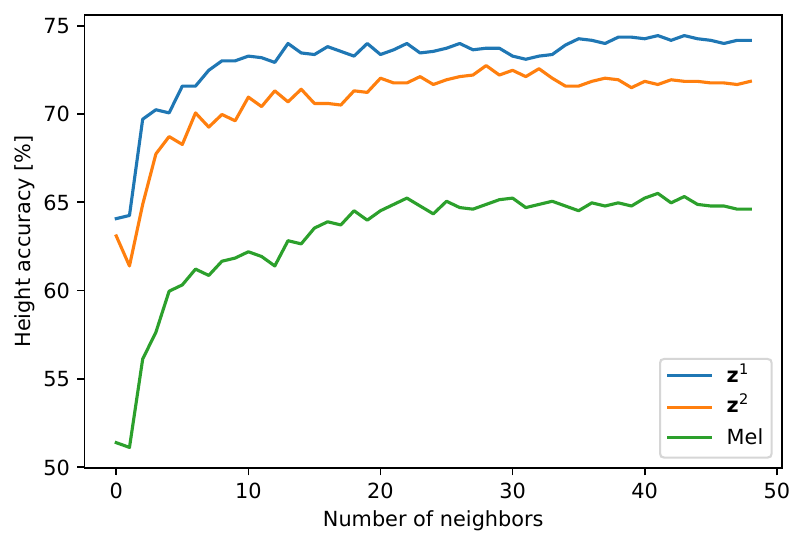}
    \caption{
    Leave-one-out $k$-nearest-neighbor accuracy with different $k$ for
    (a) the speaker's gender and
    (b) the height of male speakers (female speakers yield a similar result).
    }
    \label{fig: knn fraction latent space height and gender}
\end{figure*}

We visualize the performance of a $k$-nearest-neighbour classifier for classification of speaker gender and height in \figref{fig: knn fraction latent space height and gender}.
The classifier is fitted to time-averaged latent representations and Mel-features. 
We divide the height into three classes: below $\SI{175}{cm}$, above $\SI{185}{cm}$ and in-between. Compared to phonemes, the gender and height of a speaker are global attributes that affect the entire signal. 
In both cases, we see improved performance from using the learned latent space over Mel-features. Notably, $\vz^2$ is outperformed by the Mel-features for gender identification which may indicate that $\vz^2$ learns to ignore this attribute compared to $\vz^1$.

\begin{figure}[t!]
     \centering
     \hfill
     \begin{subfigure}[b]{0.48\textwidth}
         \centering
         \includegraphics[width=\textwidth]{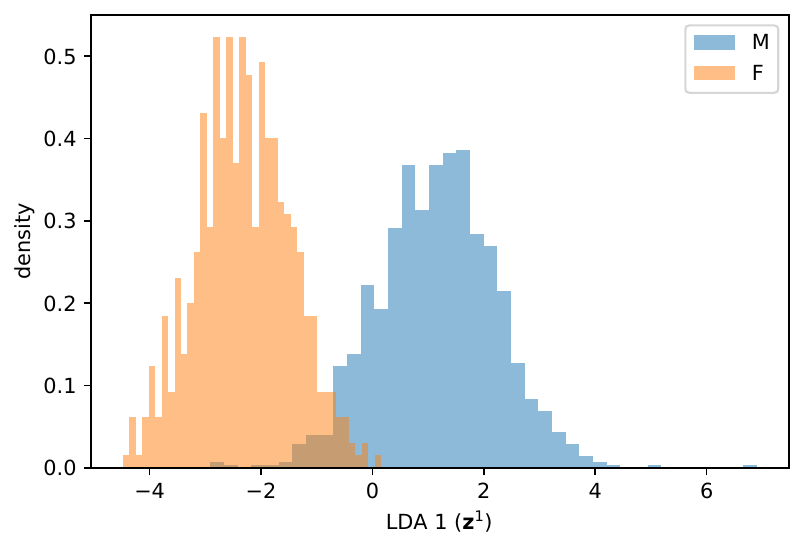}
         \caption{}
         \label{fig: latent space gender}
     \end{subfigure}
     \hfill
     \begin{subfigure}[b]{0.48\textwidth}
         \centering
         \includegraphics[width=\textwidth]{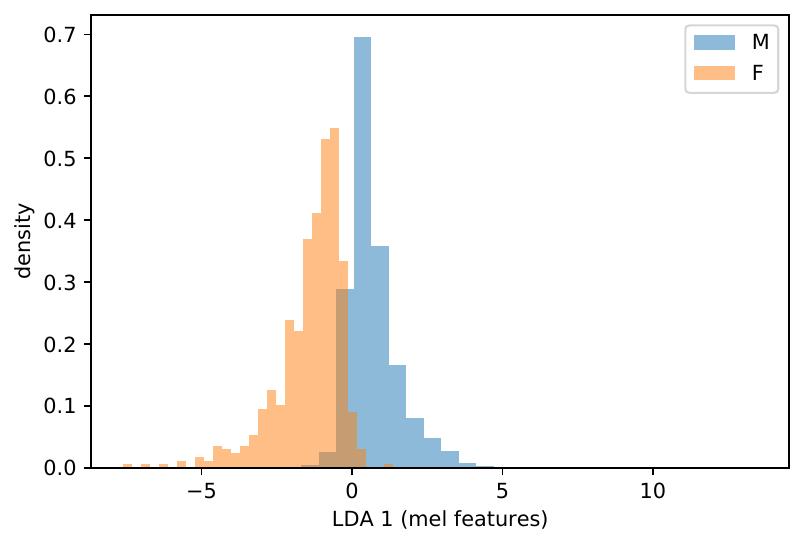}
         \caption{}
         \label{fig: mel features gender}
     \end{subfigure}
    \caption{Clustering of speaker gender in an one-dimensional linear subspace defined by a linear discriminant analysis of the CW-VAE latent space and of a time-averaged mel spectrogram. The total overlap is slightly smaller in the subspace of the CW-VAE latent space and the separation between the distribution peaks is larger.}
    \label{fig: latent space visualization gender}
\end{figure}

\begin{figure}[t!]
     \centering
     \hfill
     \begin{subfigure}[b]{0.48\textwidth}
         \centering
         \includegraphics[width=\textwidth]{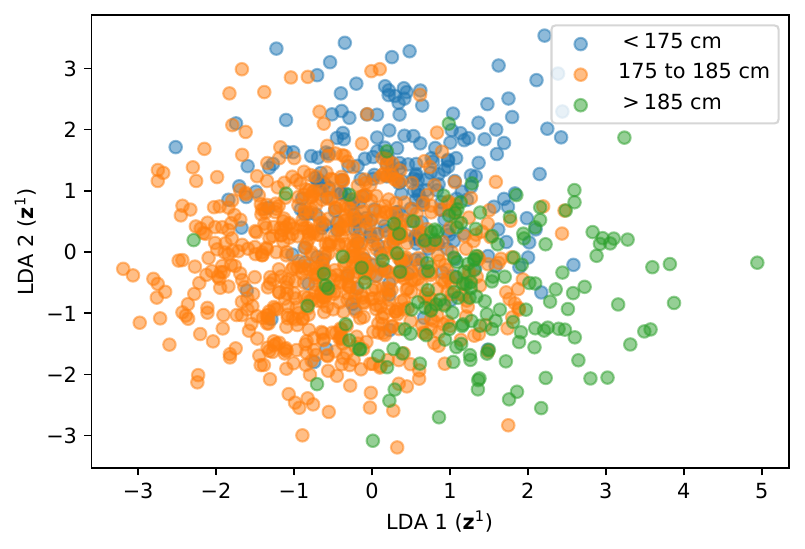}
         \caption{}
         \label{fig: cwvae latent z1 height}
     \end{subfigure}
     \hfill
     \begin{subfigure}[b]{0.48\textwidth}
         \centering
         \includegraphics[width=\textwidth]{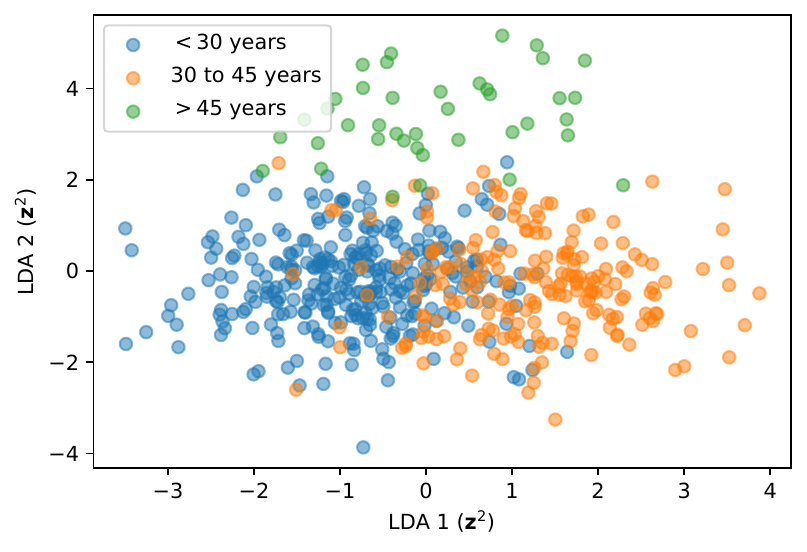}
         \caption{}
         \label{fig: cwvae latent z0 age}
     \end{subfigure}
    \caption{(a) Clustering of speaker height for male speakers and (b) speaker age for female speakers in an two-dimensional linear subspace defined by a linear discriminant analysis of the CW-VAE latent space.}
    \label{fig: cwvae latent height}
\end{figure}

We provide some additional latent space clustering of speaker gender in \figref{fig: latent space visualization gender} and of speaker height in \figref{fig: cwvae latent height}.

All results presented here are obtained with a 2-layered CW-VAE trained on $\mu$-law encoded PCM similar to the one in \tabref{tab: likelihoods timit}.


\section{Distribution of phoneme duration in TIMIT}\label{app: timit phoneme distributions}
\begin{figure}[t!]
    \centering
    \hfill
    \begin{subfigure}[b]{\textwidth}
        \centering
        \includegraphics[width=\textwidth]{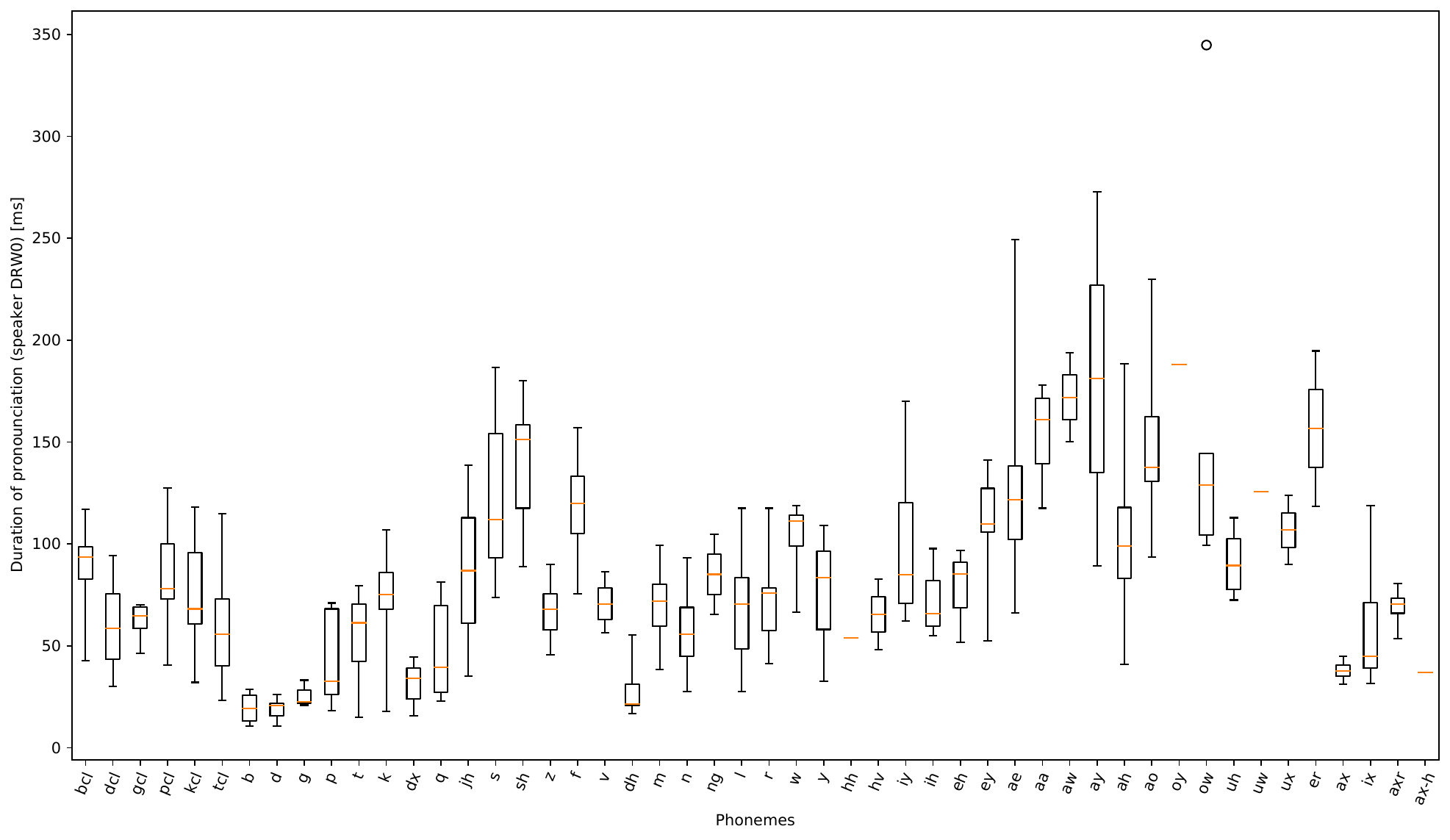}
        \caption{}
        \label{fig: timit validation set phoneme duration DRW0}
    \end{subfigure}
    \begin{subfigure}[b]{\textwidth}
        \centering
        \includegraphics[width=\textwidth]{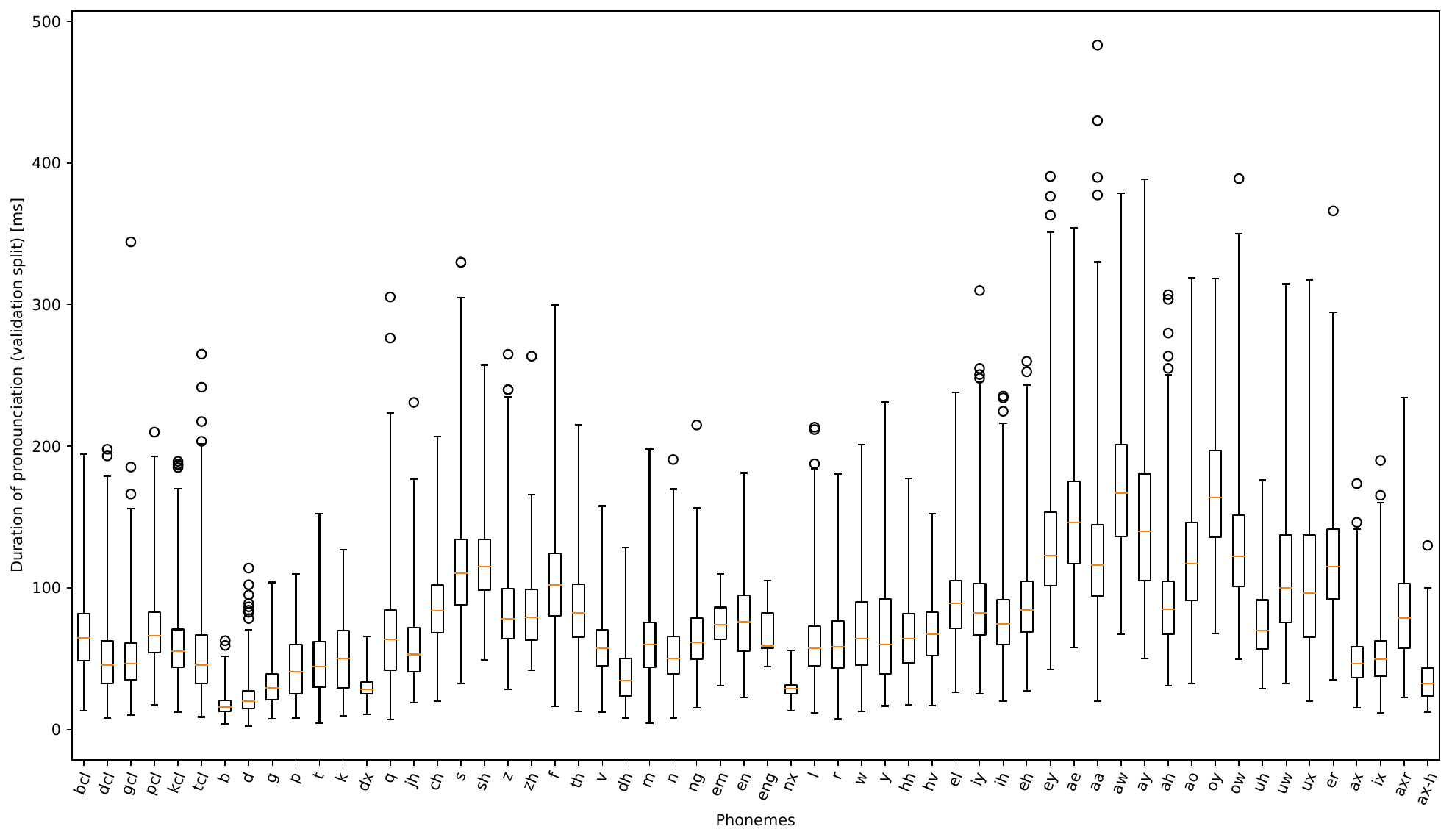}
        \caption{}
        \label{fig: timit validation set phoneme duration global}
    \end{subfigure}
    \caption{Boxplots of the duration of the pronunciation of phonemes in TIMIT for a specific speaker DRW0 in (\subref{fig: timit validation set phoneme duration DRW0}) and globally in (\subref{fig: timit validation set phoneme duration global}). Not all phonemes are pronounced by speaker DRW0 over the course of their 10 test set sentences and hence they are missing from the x-axis compared to the global durations.}
    \label{fig: timit validation set phoneme duration}
\end{figure}

In \figref{fig: timit validation set phoneme duration} we plot a boxplots of the duration of each phoneme in the TIMIT dataset. We do this globally as well as for a single speaker to show that phoneme duration can vary between individual speakers. 

A description of the phonemes used for the TIMIT dataset can be found at \url{https://catalog.ldc.upenn.edu/docs/LDC93S1/PHONCODE.TXT}.

\section{Model samples and reconstructions}\label{app: model samples and reconstructions}
We provide samples and reconstructions for some of the models considered here at the following URL: \url{https://doi.org/10.5281/zenodo.5911899}.
The samples are generated from the prior of Clockwork VAE, SRNN and VRNN and from a WaveNet by conditioning on pure zeros. All models are configured as those reported in \tabref{tab: likelihoods timit}.
Importantly, the samples are unconditional. Hence they are \emph{not} reconstructions inferred from a given input nor are they conditioned on any auxiliary data like text.

Although sample quality is a somewhat subjective matter, we find the quality of the unconditional Clockwork VAE to be better than those of our VRNN and SRNN. WaveNet is known to produce samples with intelligible speech when conditioned on e.g. text, but unconditional samples from WaveNet lack semantic content such as words as do VRNN, SRNN and Clockwork VAE.